\documentclass[aps,pre,epsf,twocolumn,floatfix]{revtex4-2}
\usepackage[greek,english]{babel}
\usepackage{float}
\usepackage{teubner}
\usepackage{amsmath}
\usepackage{mathrsfs}
\usepackage{amssymb}
\usepackage{epsfig}
\usepackage[T1]{fontenc}
\usepackage[utf8]{inputenc}
\usepackage{color}
\usepackage[normalem]{ulem}
\usepackage{tikz}
\usepackage{xcolor}
\usepackage{blindtext}

\usetikzlibrary{positioning}

\usepackage{graphicx}
\usepackage{dcolumn}
\usepackage{bm}
\usepackage{hyperref}

\DeclareUnicodeCharacter{2061}{}

\begin{document}

\title{Critical and tricritical behavior of the $d=3$ Blume-Capel model: \\ Results from small-scale Monte Carlo simulations}

\author{Leïla Moueddene}
  \email{leila.moueddene@univ-lorraine.fr}
\affiliation{Laboratoire de Physique et Chimie Théoriques, CNRS - Université de Lorraine, UMR 7019,
Nancy, France}

\collaboration{L4 collaboration, Leipzig-Lorraine-Lviv-Coventry, Europe}

\author{Nikolaos G. Fytas}
\email{nikolaos.fytas@essex.ac.uk}
\affiliation{School of Mathematics, Statistics and Actuarial Science, University of Essex, Colchester CO4 3SQ, United Kingdom}

\author{Bertrand Berche}
 \email{bertrand.berche@univ-lorraine.fr}
\affiliation{Laboratoire de Physique et Chimie Théoriques, CNRS - Université de Lorraine, UMR 7019,
Nancy, France}

\collaboration{L4 collaboration, Leipzig-Lorraine-Lviv-Coventry, Europe}

\date{\today}

\begin{abstract}
We investigate the location of the critical and tricritical points of the three-dimensional Blume-Capel model by analyzing the behavior of the first Lee-Yang zero, the density of partition function zeros, and higher-order cumulants of the magnetization. Our analysis is conducted through Monte Carlo simulations, intentionally using only small system sizes. We demonstrate that this approach yields excellent results for studying the critical behavior of the model. Our findings indicate that at the tricritical point, where logarithmic corrections are anticipated, the numerical results align closely with the theoretical exponents describing these corrections. These expected values are then employed to accurately determine the coordinates of the tricritical point. At the model's critical point, the corrections correspond to those of the three-dimensional Ising model criticality, which we also use to precisely ascertain the critical temperature at zero crystal field. Additionally, we utilize more traditional thermodynamic quantities to validate the self-consistency of our analysis.
\end{abstract}

\maketitle

\section{Introduction}
\label{sec:intro}

Leading  singularities of thermodynamic quantities  exhibiting critical behavior at a second-order phase transition are characterized by power laws  in the reduced temperature $t$ and external field $h$. In the thermodynamic limit, at $h=0$, the correlation length behaves as $\xi_{\infty} (t)  \sim |t|^{-\nu}$, and the specific heat, susceptibility, and low-temperature magnetization respectively as $C_{\infty} (t) \sim |t|^{-\alpha}$,  $\chi_{\infty} (t) \sim |t|^{-\gamma}$,  and  $m_{\infty} (t) \sim |t|^{\beta}$ (for $t < 0$). The exponents that appear in these power laws define a universality class and identical behaviors are expected for different systems in the same universality class, usually determined by general properties like ground-state symmetries, space dimensionality, and interaction range.  

However, at the upper critical dimension -- where mean-field behavior prevails -- it is well known that besides the leading singularities, multiplicative logarithmic corrections emerge. The power laws are then modified and logarithmic-corrections exponents are expected
\begin{equation}
\begin{aligned}
\xi_{\infty} (t) \sim |t|^{-\nu} | \ln{|t||^{\hat{\nu}}}\;\; ; \;\;\;\; 
C_{\infty} (t) \sim |t|^{-\alpha} | \ln{|t||^{\hat{\alpha}}}\\
\chi_{\infty} (t) \sim |t|^{-\gamma} | \ln{|t||^{\hat{\gamma}}}\;\; ;  \;\;\;\;
 m_{\infty} (t) \sim |t|^{\beta} | \ln{|t||^{\hat{\beta}}} \text{   for  $t < 0$},  
\end{aligned}
\end{equation}
where the hatted exponents are universal and adhere to specific scaling relations~\cite{kenna_self-consistent_2006, kenna_scaling_2006}. 
Beyond the leading singularities, additive corrections to scaling also appear to describe the approach to criticality along a second-order transition line and, denoting as $\omega$ the leading corrections-to-scaling exponent, an extension of the previous power laws holds, e.g., $\xi_\infty(t)\sim a|t|^{-\nu}(1+b|t|^{\nu\omega})$.

The presence of logarithmic corrections has already been studied numerically through Lee-Yang zeros, first by Kenna and Lang for the four-dimensional (4D) Ising model~\cite{kenna_renormalization_1993} and from thermodynamic quantities by Luijten~\cite{Luijten_interaction_1997}, while then revisited by Ruiz-Lorenzo using numerical simulations and very large system sizes~\cite{Ruiz-Lorenzo:2024jwf}. Below the upper critical dimension logarithmic corrections appear in certain models as well. For example the Lee-Yang zeros of the 2D XY model have been studied by Kenna and Irving~\cite{kenna_logarithmic_1995, irving_critical_1996}, as well as from Hong~\cite{hong_tensor_2022} employing tensor network calculations, but also from standard thermodynamic observables, such as the magnetic susceptibility by Janke ~\cite{janke_logarithmic_1997} and Hasenbusch~\cite{hasenbusch_two_2005}.  

In a recent paper dedicated to the memory of our friend Ralph Kenna, a pioneer in this field of Statistical Physics~\cite{moueddene_ralph_2024}, two of us have revisited the derivation of scaling laws among the exponents describing logarithmic corrections. There, to illustrate the purpose, a preliminary study of the 3D Blume-Capel model~\cite{Blume,CAPEL} at its tricritical point was undertaken. As was also highlighted in Ref.~\cite{moueddene_ralph_2024}, the Lee-Yang zeros are of particular interest due to the fact that their finite-size scaling properties capture precisely the universal aspects of the underlying phase transitions. We note here that the study of the zeros of the partition function was initially developed by Lee and Yang~\cite{Lee,Yang} who investigated the lattice gas model using the grand partition function expressed as a polynomial with a complex fugacity. As it is well-known the zeros of the partition function are associated with singularities in the free energy and are defined in the complex plane of the external magnetic field.  

Building on the preliminary results presented in Ref.~\cite{moueddene_ralph_2024}, we propose a method to investigate the phase diagram of the 3D Blume-Capel model. Our approach aims to refine the transition line by analyzing the finite-size scaling behavior of Lee-Yang zeros and high-order cumulants of the magnetization, while taking into account corrections to scaling. To the best of our knowledge, there has been no dedicated numerical analysis of the tricritical exponents in the Blume-Capel model that incorporates logarithmic corrections. The availability of both analytical and highly accurate numerical values for the leading exponents, as well as for those related to the corrections-to-scaling at the tricritical point and along the second-order transition line (Ising universality class), enables us to impose stringent constraints on the phase diagram. In particular we study the model at (i) its zero crystal-field critical point and (ii) exactly at the tricritical point through hybrid Monte-Carlo simulations that combine in an optimized manner the Wolff cluster update with the Metropolis single spin-flip dynamics. For the determination of the first zero we use the cumulant method, originally developed by Deger \emph{et al.}, and successfully tested in various models~\cite{Deger_2020WEISS,Deger_2018, Deger_2019,Deger_2020ISING,moueddene_critical_2024}.

A notable aspect of this work is that the techniques we employ are robust enough to yield reliable and accurate results even from simulations of small system sizes, especially by today’s standards. Given the growing concern about the environmental impact of scientific research, particularly the significant carbon footprint associated with large-scale numerical simulations~\cite{astro}, it is becoming increasingly necessary to justify the use of extensive computational resources. Our approach prioritizes \emph{small-scale desktop simulations}, offering a more sustainable alternative.

The remainder of the paper is structured as follows: In the next Sec.~\ref{sec:simulations} we introduce the Blume-Capel model and outline the specifics of our simulation protocol, including the relevant observables that will support our finite-size scaling analysis. The numerical results and analysis are presented in Sec.~\ref{sec:results}. Finally, Sec.~\ref{sec:conclusions} summarizes our key findings and concludes the paper.

\section{Simulation framework}
\label{sec:simulations} 

\subsection{Blume-Capel model}
\label{subsec:model}

The Blume-Capel model is a specific case of the more general Blume-Emery-Griffiths model~\cite{PhysRevA.4.1071}. This family of models has greatly contributed to the study of tricriticality in condensed-matter physics. It is also a model of great experimental interest since it can describe many physical
systems, including among others, liquid crystals and $^3$He-$^4$He mixtures \cite{PhysRevA.4.1071}. The Hamiltonian of the spin-$1$ Blume-Capel model reads as~\cite{Blume,CAPEL}
\begin{equation}
\label{eq:hamiltonian}
    \mathcal{H}=-J\sum_{\langle i,j\rangle}\sigma_i \sigma_j + \Delta \sum_i \sigma_i^2 - H\sum_i \sigma_i,
\end{equation}
where the spin variables take on the values $\sigma_i = \{-1, 0, +1\}$, $J >0$ is the ferromagnetic exchange interaction, $\Delta$ denotes the crystal-field coupling that controls the density of vacancies ($\sigma_i = 0$), and $H$ defines the external magnetic field. Of course for $\Delta = -\infty$, vacancies are suppressed and the Hamiltonian~(\ref{eq:hamiltonian}) reduces to that of the Ising ferromagnet. 

A reproduction of the phase diagram of the Blume-Capel model in the ($\Delta$, $T$)-plane can be found in many papers, see for example Ref.~\cite{Zierenberg} and references therein. For $H = 0$, a quantitative description of the phase diagram is as follows: For  small $\Delta$ there is a line of continuous transitions
between the ferromagnetic and paramagnetic phases that crosses the vertical axis at $\Delta_{\rm c} = 0$. For large $\Delta$, on the other hand, the transition becomes discontinuous and it meets the $T = 0$ line at $\Delta_{0} = zJ/2$, where $z$ is the coordination number (here $z = 6$ for the simple cubic lattice and we set $J = 1$ and $k_{\rm B} = 1$ to fix the energy and temperature scale). The two line segments meet at a tricritical point ($\Delta_{\rm t}$, $T_{\rm t}$). Narrowing down the discussion to the three-dimensional model, the transition line has been studied in several numerical studies~\cite{kutlu_tricritical_2005,PhysRevB.82.174433,fytas_universality_2013,zierenberg_parallel_2015}. In particular, the zero crystal-field critical point has been determined by Fytas and Theodorakis as $(\Delta_{\rm c}, T_{\rm c}) = (0, 3.1952(8))$~\cite{fytas_universality_2013}, and the location of the tricritical point by Deserno, $(\Delta_{\rm t}, T_{\rm t}) = (2.84479(30), 1.4182(55))$~\cite{deserno_tricriticality_1997}. These two points are of special interest for the current work also, since in three dimensions the second-order transition line lies in the ordinary Ising universality class while at the ending point we expect a mean-field transition with its logarithmic corrections~\cite{doi:10.1142/9789814417891_0001}. In fact, it is well established that the tricritical point belongs to the universality class of the $\phi^6$ theory, where the upper critical dimension is $d_{\rm u} = 3$. The logarithmic corrections in three dimensions for the tricritical point have been determined analytically by Stephen, Abrahams, and Straley~\cite{PhysRevB.12.256} and by Lawrie and Sarbach~\cite{Lawrie}. One of our goals in this work is to validate and potentially refine these values while reducing the computational effort required.

Closing this section, we should note that the very early experimental studies on $^3$He-$^4$He mixtures~\cite{riedel_thermodynamic_1976} have validated the tricritical approach by reproducing the expected leading exponents, although without logarithmic corrections. Additionally, experimental investigations of the uniform (nonordering) magnetization and heat capacity of Fe${\mathrm{Cl}}_{2}$~\cite{PhysRevB.22.4401} measured
simultaneously in the vicinity of the tricritical point indicated that the deviations of the exponents from their classical values are the result of logarithmic correction factors, aligning with the corrections predicted in Refs.~\cite{PhysRevB.12.256, Lawrie}.

\subsection{Numerical details}
\label{subsec:numerics}

We employ a hybrid numerical scheme that combines effectively a Wolﬀ single-cluster update~\cite{wolff} for the $\pm 1$ spins and a single-spin-flip Metropolis
update~\cite{hybrid, malakis_universality_2012,PhysRevB.82.174433} to account for the vacancies. Our simulations are enhanced by a histogram reweighting method, which allows us to extrapolate data obtained from simulations at fixed values of the crystal field (temperature) to nearby temperature (crystal-field) ranges. Using this hybrid scheme we simulate, under periodic boundary conditions, systems with linear sizes in the range $L = 12 - 22$ (hereafter, $N = L\times L \times L$ defines the total number of spins on the lattice). Following the previously successfully tested prescription of Refs.~\cite{fytas18, Zierenberg}, in this hybrid approach an elementary Monte Carlo step is defined as $L$ Wolff steps and $3N$ Metropolis steps: more precisely, after each Wolff step, $3L$ Metropolis spin flips are carried out. We perform $900 \times N$ Monte Carlo steps per spin to ensure equilibration, followed by $900 \times 5 \times N$ Monte Carlo steps per spin for the collection of numerical data. As already mentioned above, our simulations are carried out at the two special points along the phase boundary, the zero crystal field and the tricritical point using the literature results $(\Delta_{\rm c}, T_{\rm c}) = (0, 3.1952)$~\cite{fytas_universality_2013} and $(\Delta_{\rm t}, T_{\rm t}) =  (2.8450, 1.4182)$~\cite{deserno_tricriticality_1997}, respectively. A comment for the application of finite-size scaling to the numerical data: We restrict ourselves to data with $L \geq L_{\rm min}$. As usual, to determine an acceptable $L_{\rm min}$ we make use of the standard $\chi^2$ test of goodness of fit~\cite{press:92}. Finally, errors are computed using the Jackknife method~\cite{efronjackknife}.

\subsection{Observables}
\label{subsec:obs}

To extract the critical behavior of the model we focus on the usual thermodynamic quantities, the thermal average are denoted as $\langle O \rangle$ where $O$ is the thermodynamic quantity. The magnetic susceptibility is written 
 \begin{equation}
 \label{eq-chi}
\chi = \frac{\langle M^2\rangle -\langle |M|\rangle ^2}{k_{\rm B}T},
\end{equation}
and the magnetocaloric coefficient
 \begin{equation}
 \label{eq-m_T}
\chi_{T} = |{\langle E|M|\rangle -\langle E\rangle \langle |M|\rangle }|.
 \end{equation}
At the same time we also consider their crystal-field analogues. In particular, a specific-heat-like quantity $\chi_2$ derived from the derivatives of the partition function with respect to the crystal field $\Delta$
 \begin{equation}
 \label{eq-chi_2}
\chi_2 = \frac{\langle E_\Delta^2\rangle -\langle E_\Delta \rangle ^2}{k_{\rm B}T},
\end{equation}
where $E_\Delta=\sum_{i} \sigma_i^2$  is the crystal field energy. One can also define $\chi_{12}$, a quantity similar to the magnetocaloric coefficient, obtained exclusively from the crystal-field contribution to the energy
 \begin{equation}
 \chi_{12} = |{\langle E_\Delta|M|\rangle -\langle E_\Delta\rangle \langle |M|\rangle }|.
\label{eq-chi12}
\end{equation}

The finite-size scaling of the quantities defined in Eqs.~(\ref{eq-chi}) and (\ref{eq-chi_2}) is controlled by dominant power-laws in $\sim L^{\gamma/\nu}$ and $\sim L^{\alpha/\nu}$ respectively, while the observables defined in Eqs.~(\ref{eq-m_T}) and (\ref{eq-chi12}) both scale as $\sim L^{{(1-\beta)}/\nu}$. Other quantities of central interest in the present work are the logarithmic derivatives of the n\textsuperscript{th}-order of the magnetization with respect to the inverse temperature ($K \equiv 1/T$)
\begin{equation}
\label{eq:log_der_T}
    \frac{\partial  \ln \langle M^n \rangle }{\partial K}=   \frac{\langle M^n \mathcal{H}\rangle}{\langle M^n \rangle} -  \langle \mathcal{H} \rangle,
\end{equation}
and the crystal field $\Delta$
\begin{equation}
\label{eq:log_der_Delta}
    \frac{\partial \ln \langle M^n \rangle }{\partial \Delta}=K  \left( \langle E_{\Delta} \rangle - \frac{\langle M^n E_{\Delta} \rangle}{\langle M^n \rangle}. \right)
\end{equation}
This latter definition in Eq.~(\ref{eq:log_der_Delta}) serves as a more suitable candidate for study precisely at the tricritical point, where the critical line is nearly perpendicular to the crystal-field axis in the  $(\Delta,T)$-plane. Note that the maxima of these logarithmic derivatives of the order parameter (both with respect to $K$ and $\Delta$) are expected to scale as $\sim L^{1/\nu}$ with the system size, providing an alternative estimation for the correlation-length's critical exponent $\nu$~\cite{malakis_universality_2012}.

\begin{figure*}[ht!]
\includegraphics[height=2.6in]{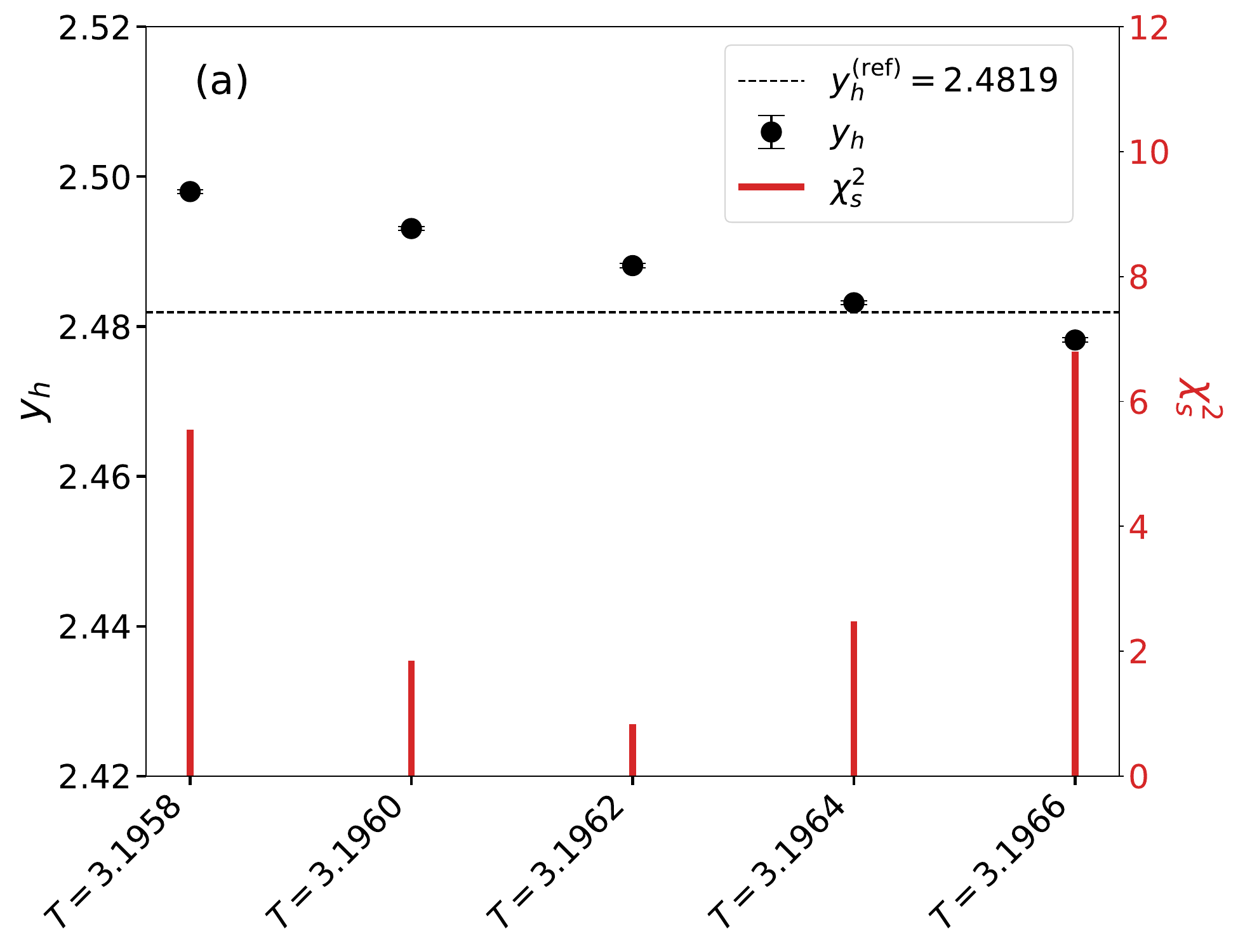}
\includegraphics[height=2.6in]{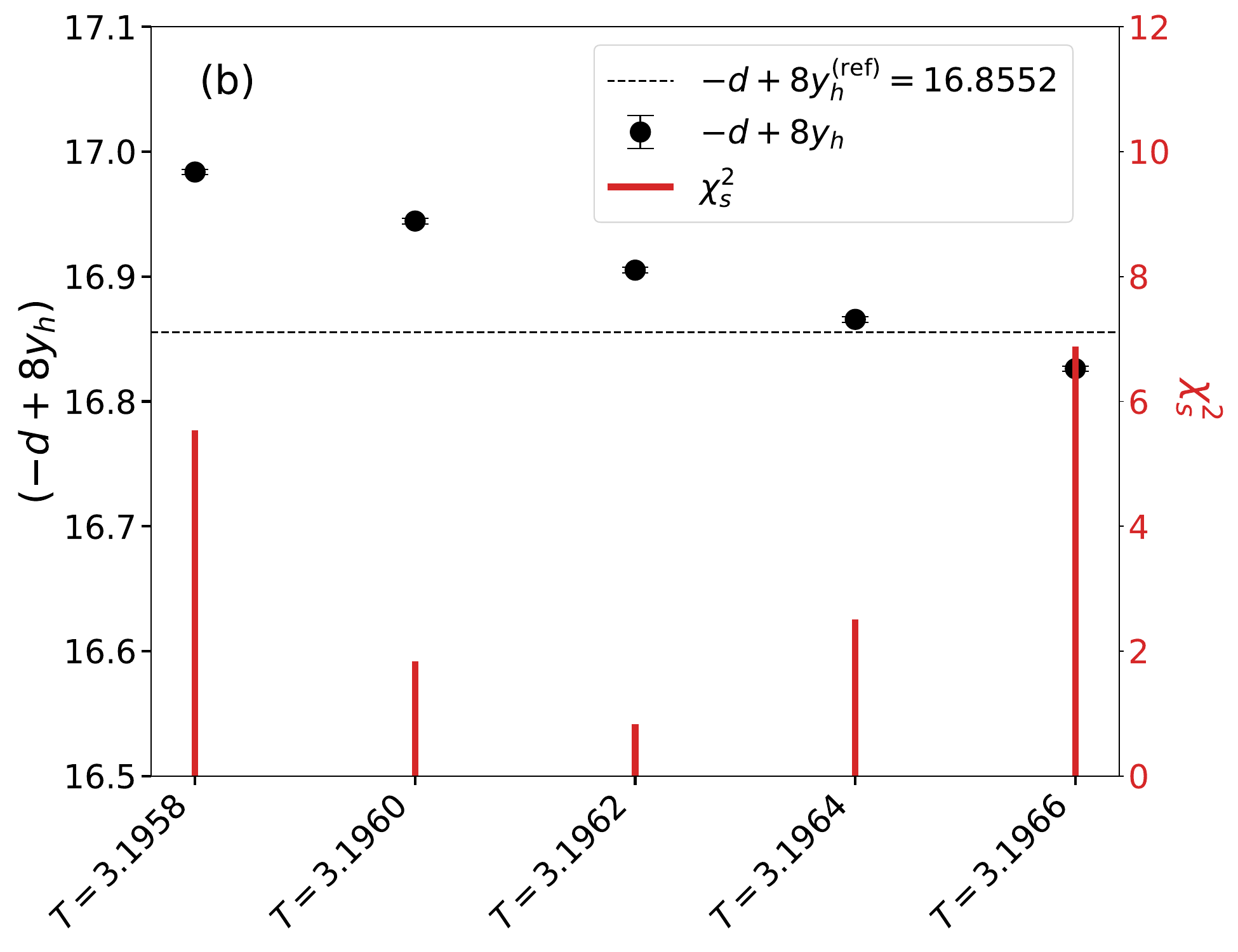}
\caption{Finite-size scaling behavior of: (a) the first Lee-Yang zero at criticality from the fit $h_0 \sim aL^{-y_h}$ and (b) the magnetization cumulant at criticality from $\langle \langle M^8 \rangle \rangle \sim aL^{8y_h-3}$, both for a set of temperatures around $T = 3.1962$. Note the value of the exponent $y_h$ marked on the left vertical scale, and the quality of the fit $\chi^2_s$ on the right vertical scale.} 
\label{fig:LY-M8-exp}
\end{figure*}

An alternative approach to studying critical phenomena involves analyzing specific cumulants of thermodynamic observables and the zeros of the partition function (since these are associated with critical singularities in the free energy). In fact as it was shown explicitly in Ref.~\cite{moueddene_critical_2024} for the $d = 2$ Blume-Capel model, the behavior of the magnetization cumulants $\langle \langle M^n \rangle \rangle$ (the double brackets refer to the cumulants) is designed for an accurate determination of the underlying critical exponents even from the study of very small system sizes. The n\textsuperscript{th} moment of the magnetization (here in the presence of a magnetic field) can be defined as the n\textsuperscript{th} derivative of the free energy with respect to that magnetic field
\begin{equation}
    \langle \langle M^n(H) \rangle \rangle = (-1)^n N^{-1}\partial_H^n \ln \mathcal{Z},
\end{equation}
where $\mathcal{Z}$ is the partition function. Another definition can be derived from the central moments 
\begin{equation}
    \langle \langle M^n(H) \rangle \rangle = \langle (M-\langle M\rangle )^n \rangle /N,
\end{equation}
whose finite-size scaling follows the Ansätze $\langle \langle M^n(0) \rangle \rangle  \sim L^{-d+ny_h}$ where $y_h$ is the renormalization-group eigenvalue associated to the magnetic field $h$, also defined as $y_h=\beta\delta/\nu$. Returning now to the case of Lee-Yang zeros, which are the remaining quantities of interest, the first zero can be defined from the magnetization cumulants for large $n$ via~\cite{Deger_2020WEISS,Deger_2018,Deger_2019,Deger_2020ISING}
\begin{equation}
\operatorname{Im}[h_0] \approx \pm \frac{1}{K}\sqrt{ 2n(2n+1) \left|
\frac{\langle \langle M^{2n}(0) \rangle \rangle}{\langle \langle M^{2(n+1)}(0) \rangle \rangle }\right|},
\end{equation}
where $n = 3$ in the present study. 

Finally, as we are also interested in improving the description of the phase diagram we resort to the study of the density of zeros of the partition function. This is a very efficient approach that has been successfully applied to a series of models in Statistical Physics, such as the Ising ferromagnet~\cite{PhysRevE.49.5012,Ruiz-Lorenzo:2024jwf}, the Heisenberg model~\cite{Gordillo_Guerrero_2013}, the Potts, $\mathrm{SU}(3)$, and Abelian Surface Gauge models~\cite{janke_analysis_2001}. In this framework, the cumulative distribution function of the zeros is given by 
\begin{equation}
    G_L(r_j)=(2(j+1)-1)/2L^d,
\end{equation}
where $r_j$ is the $(j+1)^{\rm th}$-zero of the partition function, and $j$ labels the zeros in increasing order, starting from $j = 0$. Our study is carried out for the density of the Lee-Yang zeros for which the leading critical behavior in the thermodynamic limit is described by
\begin{equation}
G_\infty(h) \sim h^{d/y_h}.
\end{equation}
In the case of a first-order phase transition, $d/y_h = 1§$, since the value of $y_h$ is effectively replaced by $d$.

\section{Results and Analysis}
\label{sec:results}

\subsection{Critical Ising scaling behavior}
\label{sec:critical} 

The aim of this section is to investigate the critical scaling behavior of the Ising universality class at $\Delta_{\rm c} = 0$. Previous numerical studies at this critical point have provided the following estimates for the corresponding critical temperature: (i) $T_{\rm c} = 3.1952(8)$ obtained from Wang-Landau simulations on system sizes $L=8-64$ by analyzing the shift scaling behavior of various pseudocritical temperatures~\cite{fytas_universality_2013}, and (ii) $T_{\rm c} = 3.20(1)$, determined through simulations using a cellular automaton with $L=8-24$~\cite{kutlu_tricritical_2005}. 

\begin{figure*}[t!]
\includegraphics[height=2.5in]{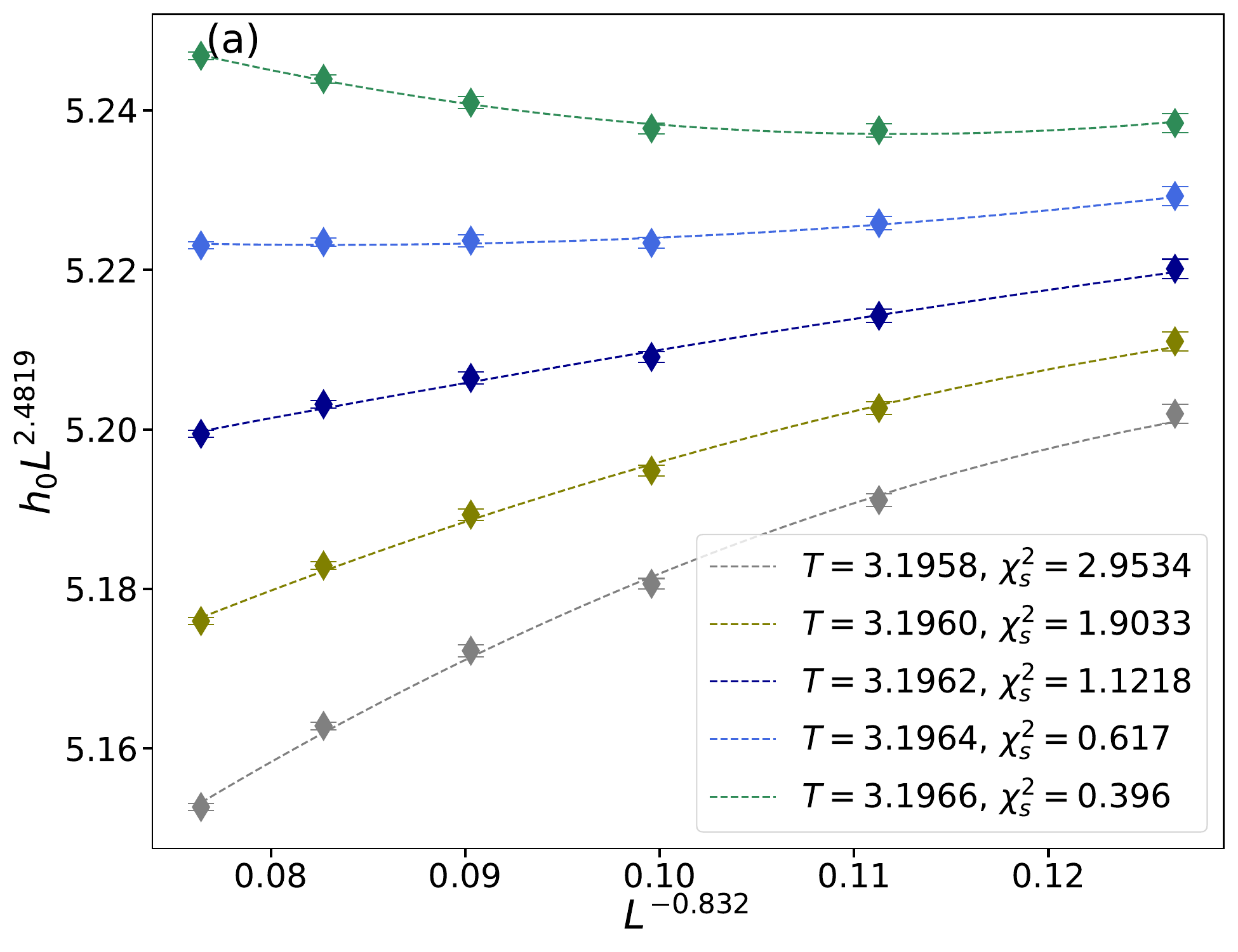}
\includegraphics[height=2.5in]{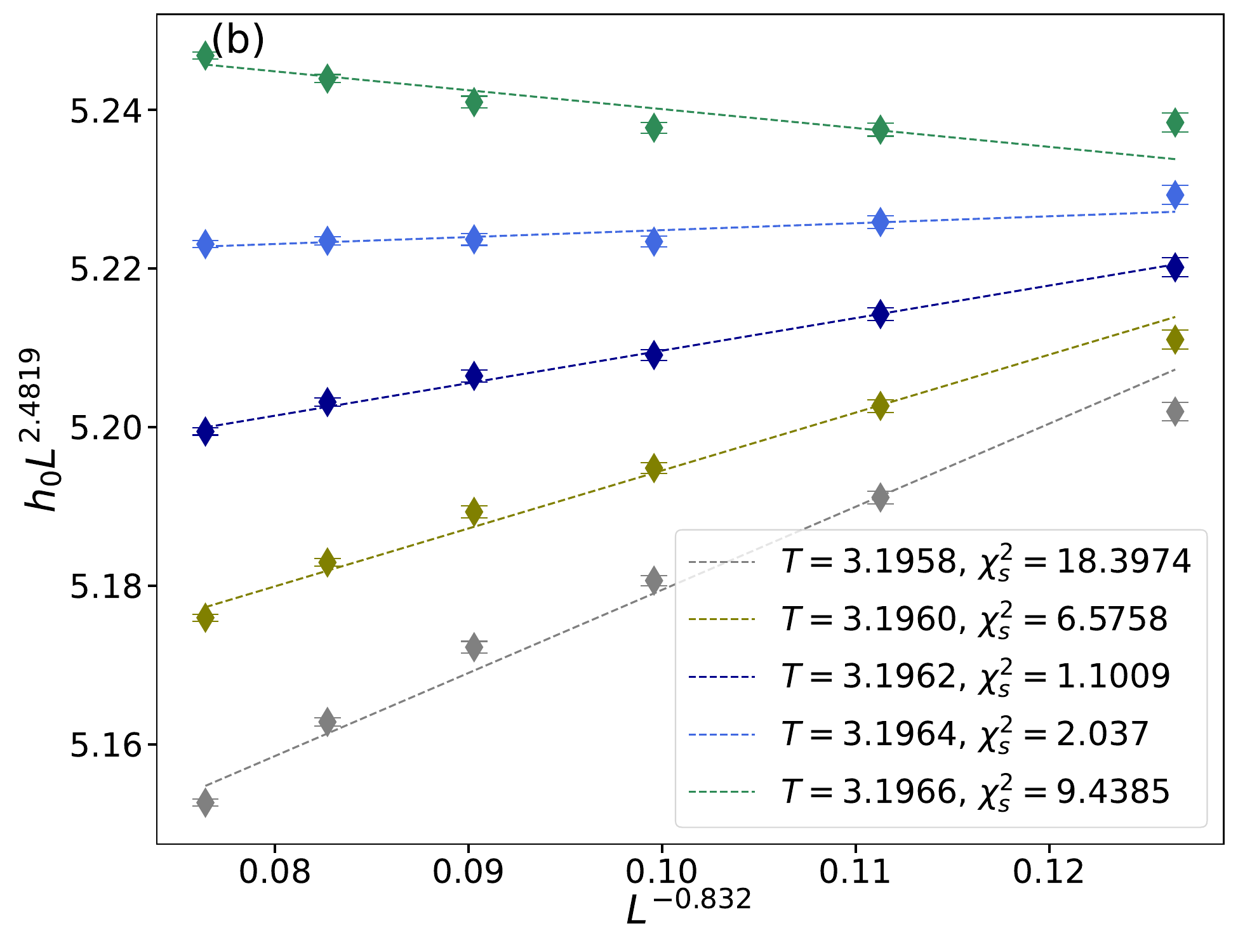}
\includegraphics[height=2.5in]{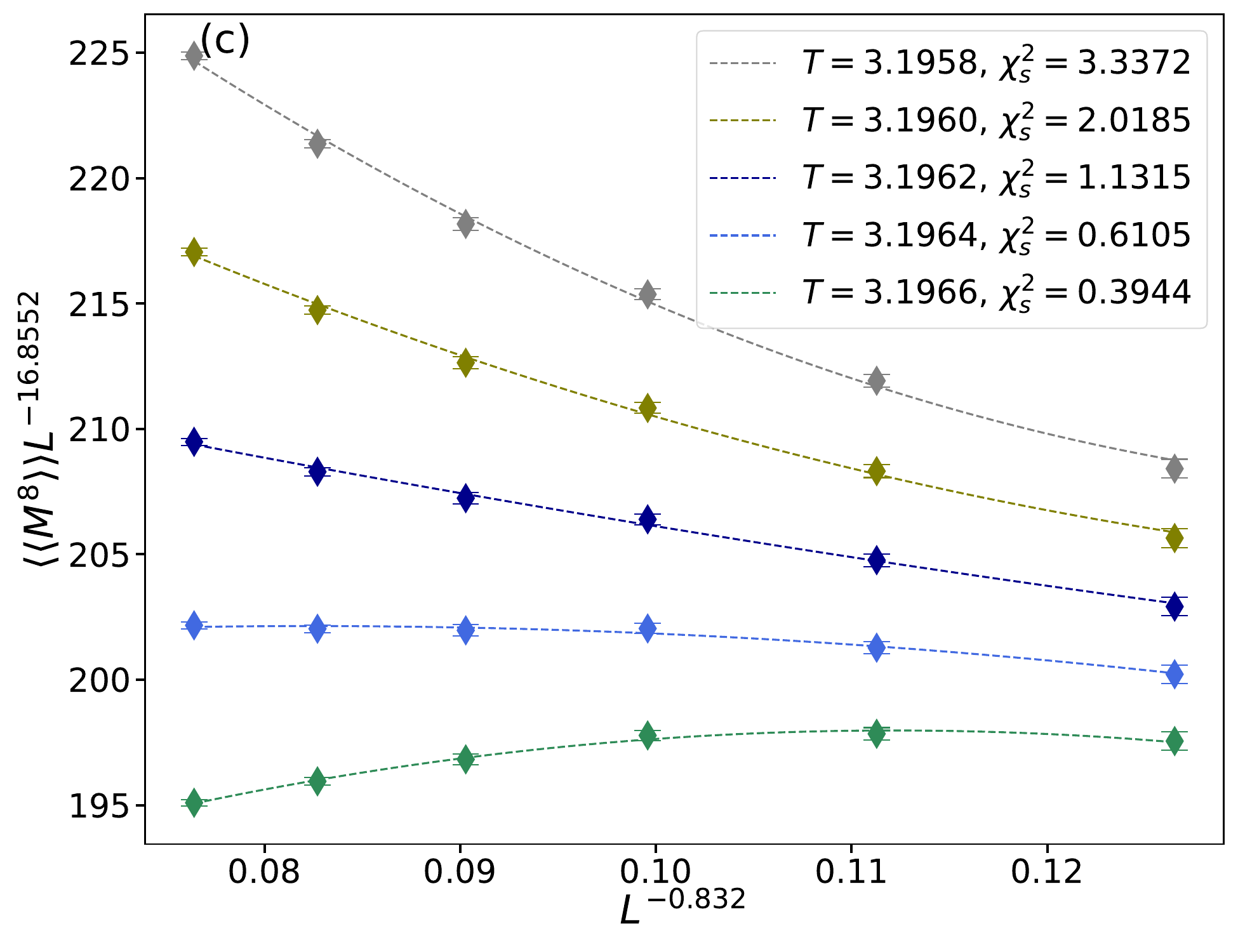}
\includegraphics[height=2.5in]{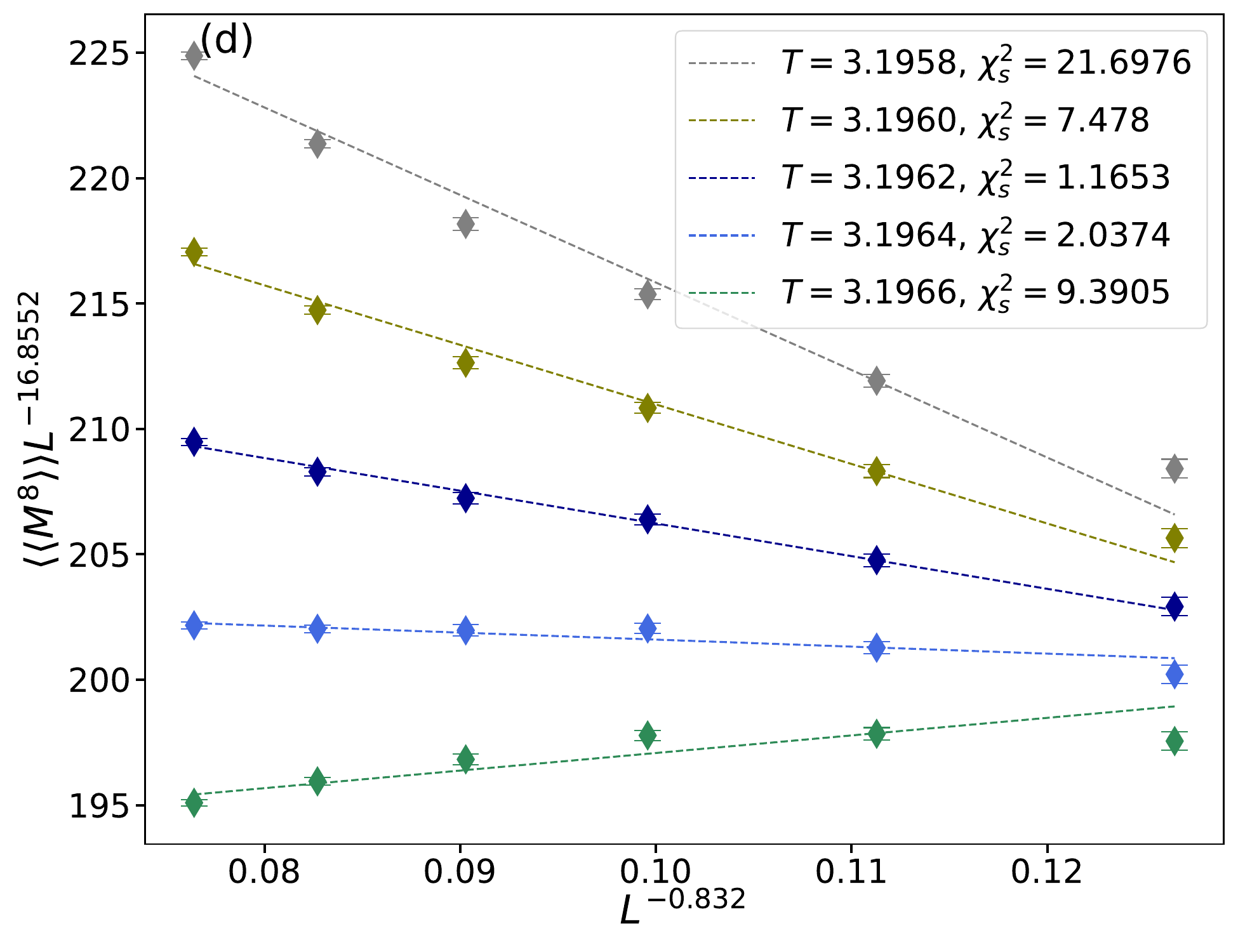}
\label{fig:M8-w}      
\caption{Finite-size scaling analysis at criticality of the first Lee-Yang zero from fits of the form $h_0 L^{2.4819}  \sim a(1 + bL^ {-0.832} + cL^{-1.168})$ [panel (a)] and $h_0 L^{2.4819} \sim a(1 + bL^ {-0.832})$ panel (b)]. Similar analysis for the $8^{\rm th}$-order magnetization cumulant using now fits of the form $\langle \langle M^8 \rangle \rangle L^{-16.8552} \sim a(1 + bL^ {-0.832} + cL^{-1.168})$ [panel (c)] and $\langle \langle M^8 \rangle \rangle L^{-16.8552} \sim a(1 + bL^ {-0.832})$ [panel (d)].} 
\label{fig:LY-w2}
\end{figure*}

Before we delve into our numerical data and analysis, we would like to remind the reader that the singular part of the free-energy density can be expressed in the standard form of a generalized homogeneous function as follows:
\begin{equation}
    \label{eq:sing}
    f^{\rm sing}=L^{-d}f^{\rm sing}(L^{y_t}t, L^{y_h}h).
\end{equation}
In the 3D Ising universality class, conformal bootstrap methods constrain the anomalous dimensions of the scaling fields to the reference values $y_t^{\rm (ref)} \approx 1.587374(4)$ and $y_h^{\rm (ref)} \approx 2.48180(14)$~\cite{Kos_2014}. These values will be utilized for comparative purposes in the following sections.

\subsubsection{Preliminary analysis}

To start, we will perform a preliminary analysis to scrutinize the reliability of our numerical results. To achieve this we determined an estimate of the exponent $y_h$ measured for different values of $T$ in the vicinity of the transition point. We therefore considered the first Lee-Yang zero, $h_0\sim aL^{-y_h}$, and the $8^{\rm th}$-order cumulant of the magnetization, $\langle \langle M^{8} \rangle \rangle\sim aL^{-d+8y_h}$ (of course the $a$-amplitudes take different values for the two quantities analyzed). A linear fit of the logarithm of $h_0$ (respectively of $\langle \langle M^{8} \rangle \rangle$) versus $\ln L$ is a straightforward task, the quality of which is assessed based on the value of the $\chi^2_s = \chi^2/$DOF, where DOF stands for the number of degrees of freedom. In both cases, the minimum of $\chi^2_s$ is obtained at $T=3.1962$ where the exponents measured are close to the expected ones. In particular, $y_h \approx 2.4881(3)$ for the Lee-Yang zeros and $(8y_h-3) \approx 16.905(3)$ for the $8^{\rm th}$-order cumulant (for which the reference value is $16.8544(11)$). These values agree with the conformal bootstrap results to an accuracy better than $99\%$. It is also instructive to correlate the values of the exponents extracted with those of the fitting quality. This is performed in Fig.~\ref{fig:LY-M8-exp} for both the first Lee-Yang zero (panel (a)) and the magnetization cumulant (panel (b)). The horizontal dashed lines represent the reference exponents, the symbols mark our numerical estimates, and the vertical lines the corresponding values of the quality of the fit. As seen in Fig.~\ref{fig:LY-M8-exp}, a strong balance is achieved between the quality of the fits and the values of the exponents. 

\subsubsection{Accurate determination of the critical temperature}

We now aim to accurately pinpoint the critical point at 
$\Delta_{\rm c} = 0$ by refining our estimate of the corresponding critical temperature $T_{\rm c}$. Given that the exponents of the 3D Ising universality class are well established, our approach will involve fixing these exponents and incorporating the known leading and subleading corrections-to-scaling, while fitting only for the amplitudes. 

We start by examining the Lee-Yang zeros, as prior research in two dimensions has shown that these quantities are highly sensitive to external parameters~\cite{moueddene_critical_2024}. For the first Lee-Yang zero we use an Ansatz of the form 
\begin{equation}
    h_0 L^{y_h} \sim a(1+bL^{-\omega}+cL^{-2\omega}),
    \label{Eq_15}
\end{equation}
where $y_h$ is fixed to its reference value, the corrections-to-scaling exponent to the value $\omega = 0.832$~\cite{PhysRevB.82.174433}, and the amplitudes $a$, $b$, and $c$ are treated as free parameters. Note that this value of $\omega$ has been numerically verified for the 3D Blume-Capel model and that at the $\Delta = 0.656(20)$ along the critical Ising line transition the amplitude of the leading corrections-to-scaling vanishes~\cite{PhysRevB.82.174433}. In Fig.~\ref{fig:LY-w2}(a) we display the fits for the parameters $a$, $b$, and $c$ at various values of the temperature in the range $T = 3.1958 - 3.1966$ (with $\Delta_{\rm c} = 0$).
\begin{figure}[ht!]
\includegraphics[width=0.48\textwidth]{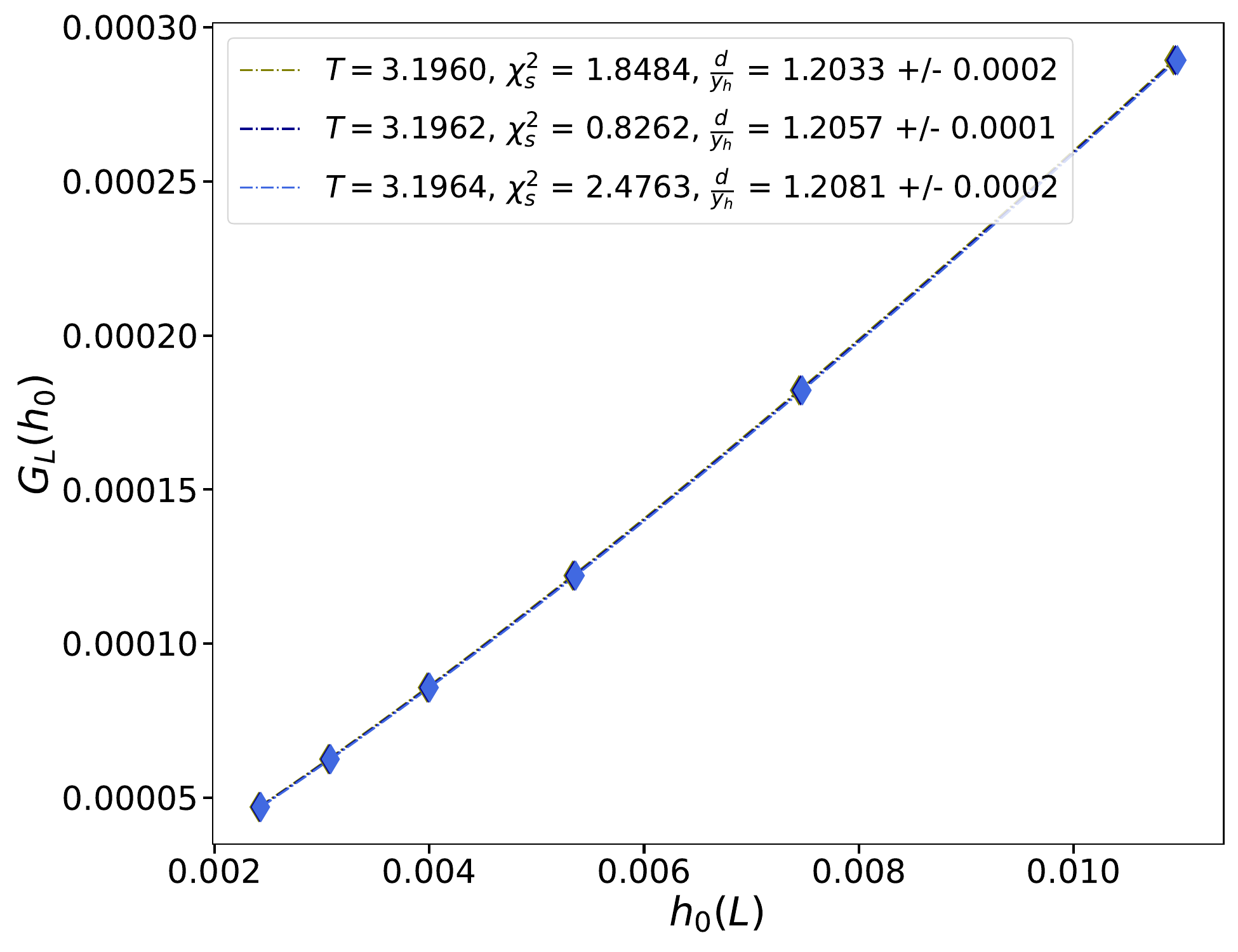}
\caption{Finite-size scaling analysis of the density of the first Lee-Yang zero close to criticality for three values of the temperature, as indicated. Note that all three curves merge in the figure and are almost superimposed.}
\label{fig:3D-density-DC}
\end{figure}
The consistency of the fits can initially be assessed graphically --  notice the straight line when $c = 0$ is fixed, along with the slight deviation from linearity in other cases. This is the case for $T = 3.1962$, where the amplitude of the coefficient $c$ is minimal in panel (a). Figure~\ref{fig:LY-w2}(b) is even more compelling, as 
$c$ is fixed to zero and the analysis is conducted without including the last correction term in Eq.~(\ref{Eq_15}). Moreover, the quality of the fits, as indicated by the $\chi^2_s$	value, seems to be optimal for the case $T = 3.1962$, reinforcing our main conclusion. In Figs.~\ref{fig:LY-w2}(c) and (d) we show a similar finite-size scaling analysis for another quantity of interest, the $8^{\rm th}$-order cumulant of the magnetization for which the expected scaling reads as
\begin{equation}
        \langle\langle M^8\rangle\rangle L^{d-8y_h} \sim a(1+bL^{-\omega}+cL^{-2\omega}).\label{Eq_16}
\end{equation}
The results are entirely consistent with those obtained from the Lee-Yang zeros, validating our estimation of the critical temperature $T_{\rm c} = 3.1962$ at $\Delta_{\rm c} = 0$. 

It is also intriguing to examine the density of zeros at this point~\cite{janke_analysis_2001}, where one expects here a second-order phase transition. Therefore the finite-size scaling behavior for the Lee-Yang zeros density is expected to be of the form 
\begin{equation}
    G_L(h_0) \sim a h_0^{d/y_h},
    \label{Eq_densityLY}
\end{equation}
where $d/y_h^{\rm (ref)} \approx 1.2088$ for the 3D Ising model. Again, the leading amplitude is simply denoted by $a$ (but it is unrelated to the previous amplitudes). Away from criticality, an additional constant should be included in the scaling form~(\ref{Eq_densityLY}). However, we first verified that this is not necessary, suggesting that we are indeed in the vicinity of the critical point.

\begin{figure}[ht!]
\includegraphics[width=0.45\textwidth]{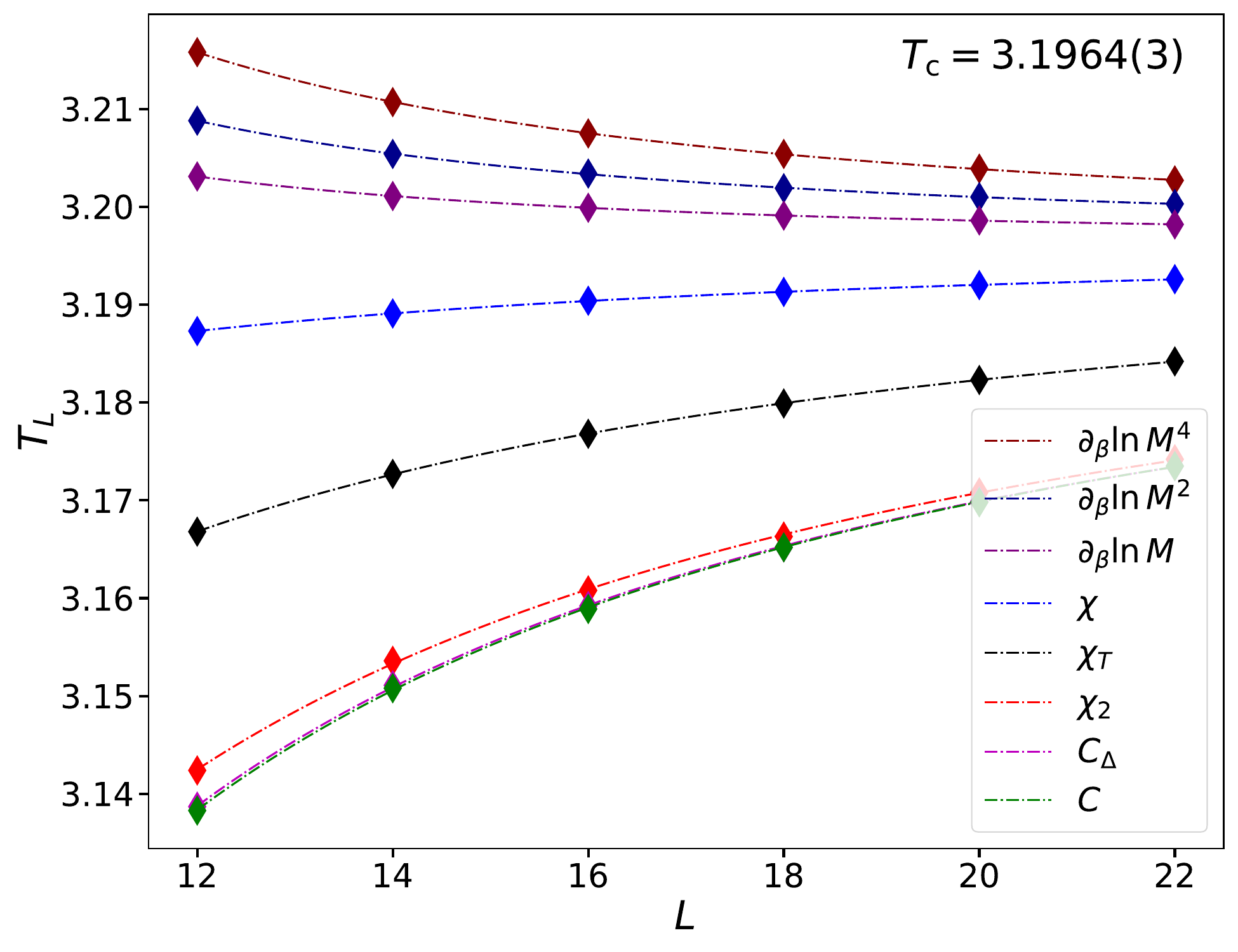}
\caption{Shift behavior of several pseudocritical temperatures $T_{L}$ defined in the main text.}
\label{fig:3D-TC-TL}
\end{figure}

In Fig.~\ref{fig:3D-density-DC}, we present the results for three different temperatures, as indicated in the panel. The curves are nearly indistinguishable, yet the best fit quality is achieved for the data corresponding to $T_{\rm c} = 3.1962$, further reinforcing the findings from the previous analysis based on the first Lee-Yang zero and the magnetization cumulant. Additionally, the measured value for the exponent, $(d/y_h) \approx 1.2057$ is in excellent agreement with the expected value, showing a deviation of less than $1\%$.

To move forward, it is now instructive to analyze more traditional quantities, namely the standard thermodynamic quantities. This will not only help us verify whether the results provide a consistent picture but also allow us to assess the advantages of our current approach. An estimate of the critical temperature is typically found by studying the crossings of the Binder cumulant (or of some other universal ratio, like $\xi/L$, where $\xi$ the correlation length) or via the shift behavior of related pseudocritical temperatures measured for the various diverging quantities (for example the location of the peaks of the specific heat or susceptibility). We choose to examine this latter route here for a set of thermodynamic observables defined above in Sec.~\ref{subsec:obs}. As one also expects corrections to scaling, the scaling law describing the shift behavior of these pseudocritical temperatures $T_{L}$ reads as
\begin{equation}
T_L \sim T_{\rm c} + aL^{-y_t}(1+bL^{-\omega}).
\label{eq-multi}
\end{equation} 
In Fig.~\ref{fig:3D-TC-TL} we display a joint fit of the form~(\ref{eq-multi}) where the exponents $y_t$ and $\omega$ were fixed during the fit to their reference values, with the constraint of a shared critical temperature $T_{\rm c}$. This protocol delivers a value $T_{\rm c} = 3.1964(3)$ consistent within error bars with the previous estimate of $3.1962$.

\begin{figure}[t] 
\includegraphics[width=0.45\textwidth]{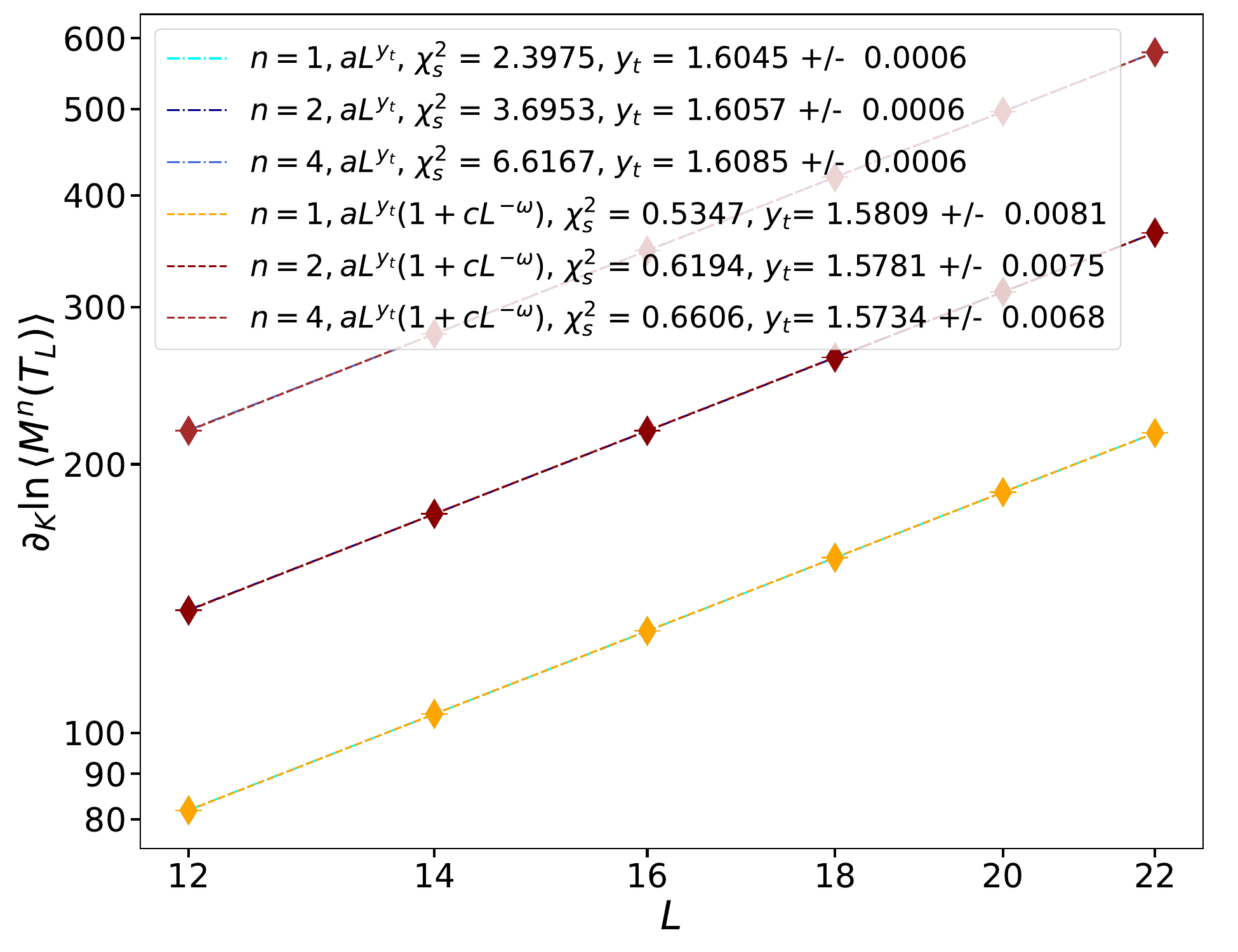}
\caption{Finite-size scaling behavior of the logarithmic derivatives of the n\textsuperscript{th}-order of the magnetization with respect to the inverse temperature at the pseudocritical point. Results for $n = 1$, $2$, and $4$ are shown. Note that for the fits including correction terms, the corrections-to-scaling exponent $\omega$ was fixed.} 
\label{fig:DC-dblogM}
\end{figure}

Finally, to ensure the self-consistency of our analysis we inspect the exponent $y_t$ at $T_{\rm c}$. One approach is via the logarithmic derivatives of the n\textsuperscript{th}-order of the magnetization with respect to the inverse temperature, defined in Eq.~(\ref{eq:log_der_T}), which scale as $\sim L^{y_t}$. Figure~\ref{fig:DC-dblogM} illustrates two distinct scaling Ans\"atze, one including corrections-to-scaling terms and the other excluding them, for $n = 1$, $2$, and $4$. The strategy employed in this paper intentionally focuses on small lattice sizes, which leads to a suboptimal fit quality for the linear models. Therefore, corrections are necessary, and by incorporating the leading corrections-to-scaling term $L^{-\omega}$ into the fit, the $\chi^2_s$ is significantly improved for all cases. Moreover, the computed exponents are close to the expected value $y_t^{\rm (ref)}\approx 1.587374(4)$, particularly for the case with $n=1$, where $y_t \approx 1.5809(81)$, resulting in a deviation of less than $0.4\%$.

\subsection{Tricritical scaling behavior}
\label{sec:tricritical} 

For the tricritical Ising universality class, $d = 3$ corresponds to the upper critical dimension $d_{\rm u}$, above which the critical exponents take on their tricritical mean-field values deduced, e.g., from a $\phi^6$ Landau expansion~\cite{10.21468/SciPostPhysLectNotes.60}. Exactly at $d = d_{\rm u}=3$, in addition to leading singularities, multiplicative logarithmic corrections also emerge.

The general scaling hypothesis for the free-energy density in the presence of logarithmic corrections can be expressed via~\cite{AktekinJSP,kenna_finite_2004}
 \begin{equation} 
f^{\rm sing}_{\rm tri}(\tau,h)=L^{-d} F (L^{y_\tau}(\ln L)^{{\hat y}_\tau}\tau, L^{y_h}(\ln L)^{{\hat y}_h}h).
    \label{eq-log}
\end{equation}
This represents a natural extension of the standard picture; however, strictly speaking, it is no longer a generalized homogeneous assumption (hence the different notations 
$f$ and $F$ for the functions on both sides of the equality). Additionally, scale invariance is broken by the presence of the logarithmic terms. Nevertheless, this form has been used extensively, particularly at the upper critical dimension of various systems~\cite{coppahat,doi:10.1142/9789814417891_0001,doi:10.1142/9789814632683_0001,kenna2024enigmaticexponentkoppastory}, and it remains prevalent in recent studies~\cite{li2024logarithmicfinitesizescalingfourdimensional}. The ``hatted'' exponents $\hat y_\tau$ and $\hat y_h$ represent the logarithmic counterparts of the anomalous scaling dimensions $y_\tau$ and $y_h$. The use of $\tau$ instead of $t$ is intended to avoid any potential confusion with tricritical notations.

Regarding the study of the Blume-Capel model, three scaling fields are needed to describe the approach to the tricritical point $(\Delta_{\rm t}, T_{\rm t})$; the even scaling fields $\tau = (T-T_{\rm t})/T_{\rm t}$, $g=(\Delta-\Delta_{\rm t})/T_{\rm t} + a\tau$, and the (odd) magnetic field $h = H/T_{\rm t}$. The scaling behavior of the free energy and its derivatives follows from the solutions of the renormalization-group equations. From the work of Refs.~\cite{PhysRevB.12.256,Lawrie}, we deduce that the free-energy density can be expressed as 
\begin{widetext}
 \begin{equation}
f^{\rm sing}_{\rm tri}(\tau,g,h)=L^{-3} \times {\mathscr F} (L^{1}(\ln L)^{\frac{4}{15}}\tau, L^{2}(\ln L)^{\frac13}g
,L^{\frac 52} (\ln L)^{\frac{1}{6}}h),
\label{eq-sws-withlogs}
\end{equation}
\end{widetext}
generalizing Eq.~(\ref{eq-log}) to a third scaling field $g$ and fixing the scaling dimensions and their hatted generalizations to their reference values
\begin{equation}
\begin{aligned}
y^{\rm (ref)}_g=2\;\; ;\quad y^{\rm (ref)}_h=\frac 52\;\; ;\quad y^{\rm (ref)}_\tau=1,\quad\\
\hat y^{\rm (ref)}_g=\frac 13\;\; ;\quad \hat y^{\rm (ref)}_h=\frac 16\;\; ;\quad \hat y^{\rm (ref)}_\tau=\frac 4{15}. 
\end{aligned}
\end{equation}

The location of the tricritical point in the $d = 3$ Blume-Capel model has been determined numerically using various methods. Deserno, employing Monte Carlo simulations in the microcanonical ensemble with system sizes ranging from $L = 8$ to $18$ the coordinates $(\Delta_{\rm t}, T_{\rm t}) = (2.84479(30), 1.4182(55))$. Zierenberg, Fytas, and Janke conducted extensive multicanonical simulations for sizes up to $L = 28$ at $T = T_{\rm t} = 1.4182$ and proposed the value $\Delta_{\rm t} = 2.8446(3)$. Their analysis was based on the shift behavior of the pseudocritical fields $\Delta_{L}$, corresponding to the locations of the peaks of a specific-heat-like quantity~\cite{zierenberg_parallel_2015}.

\begin{figure*}[ht!]
\includegraphics[height=2.5in]{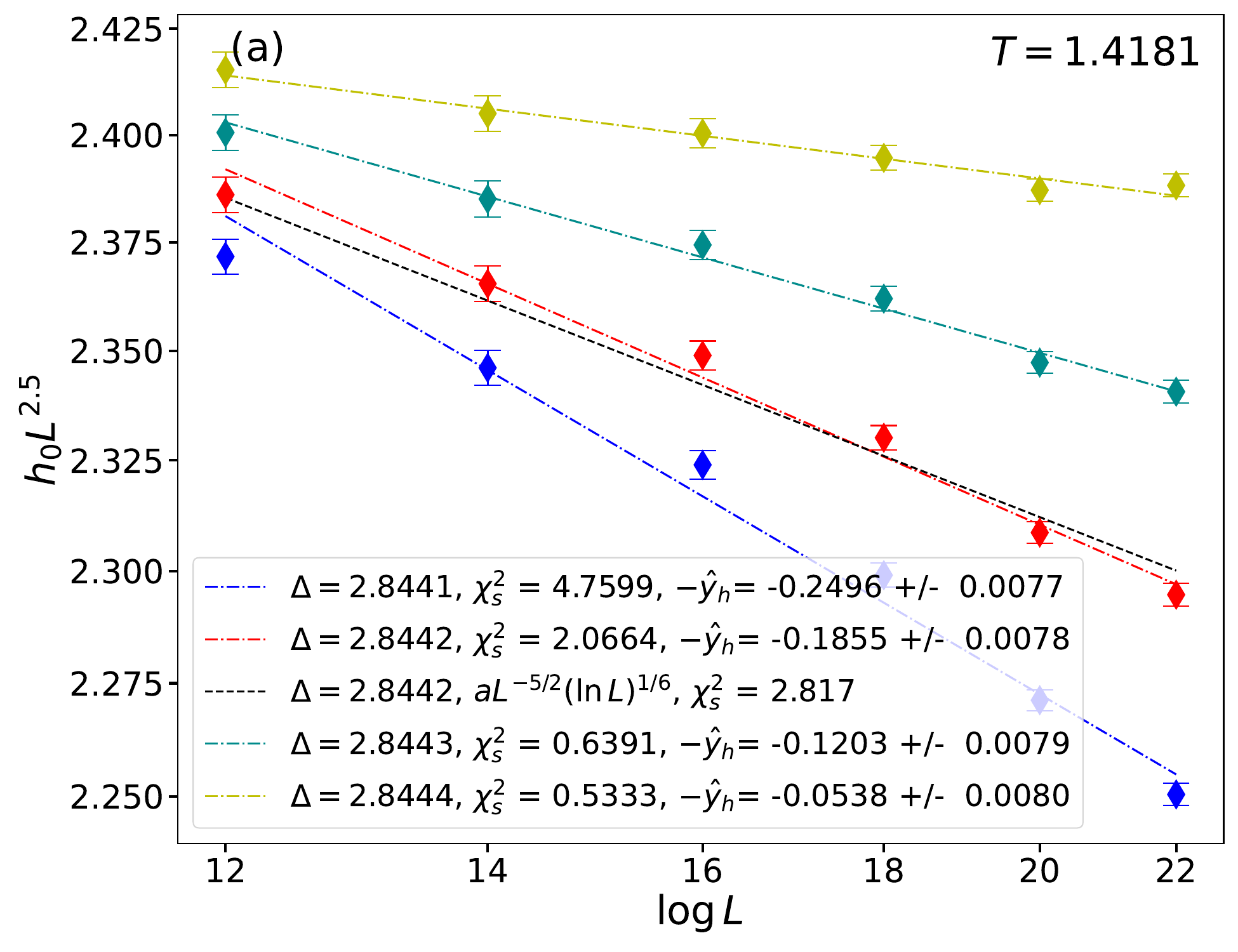}
\includegraphics[height=2.5in]{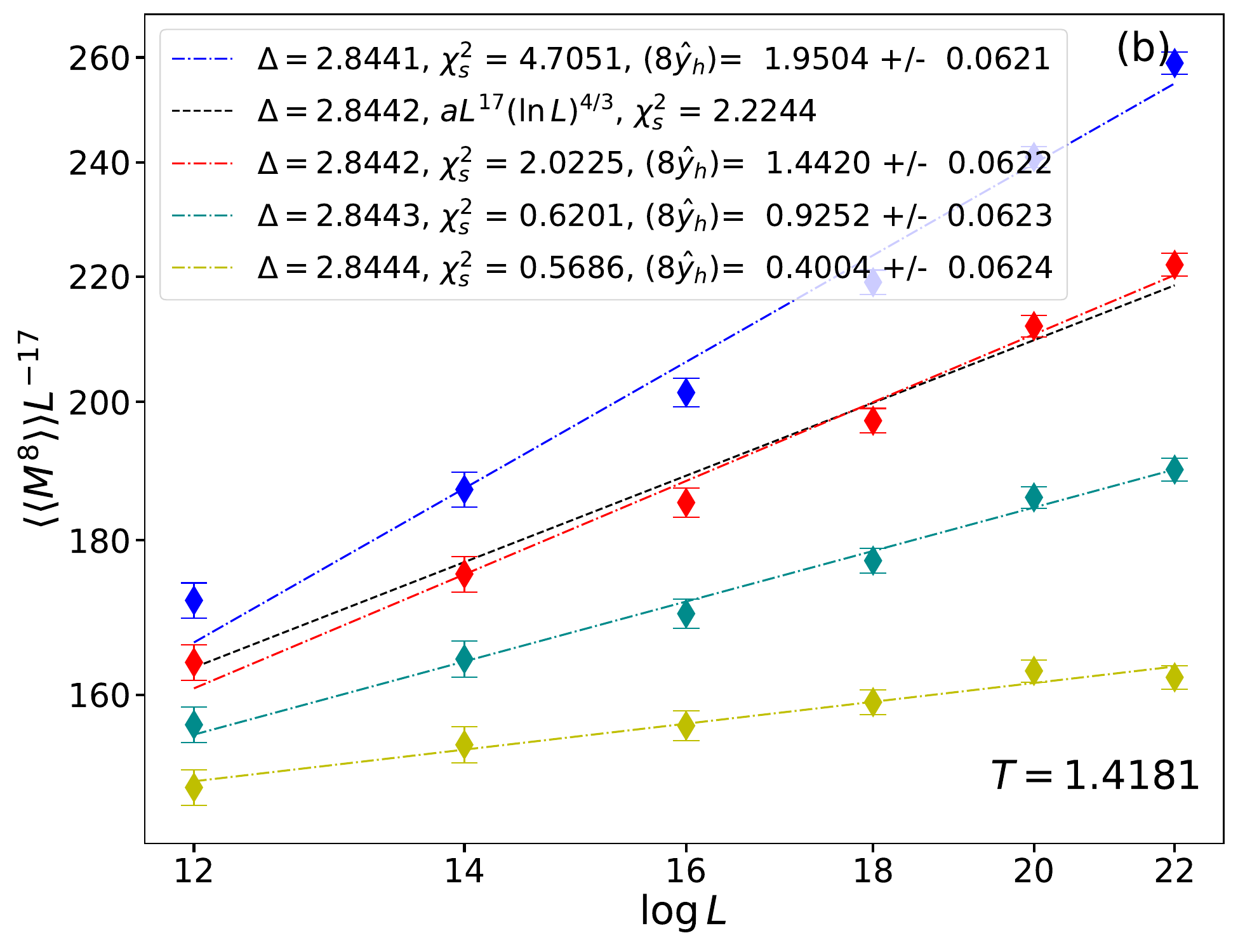}
\includegraphics[height=2.5in]{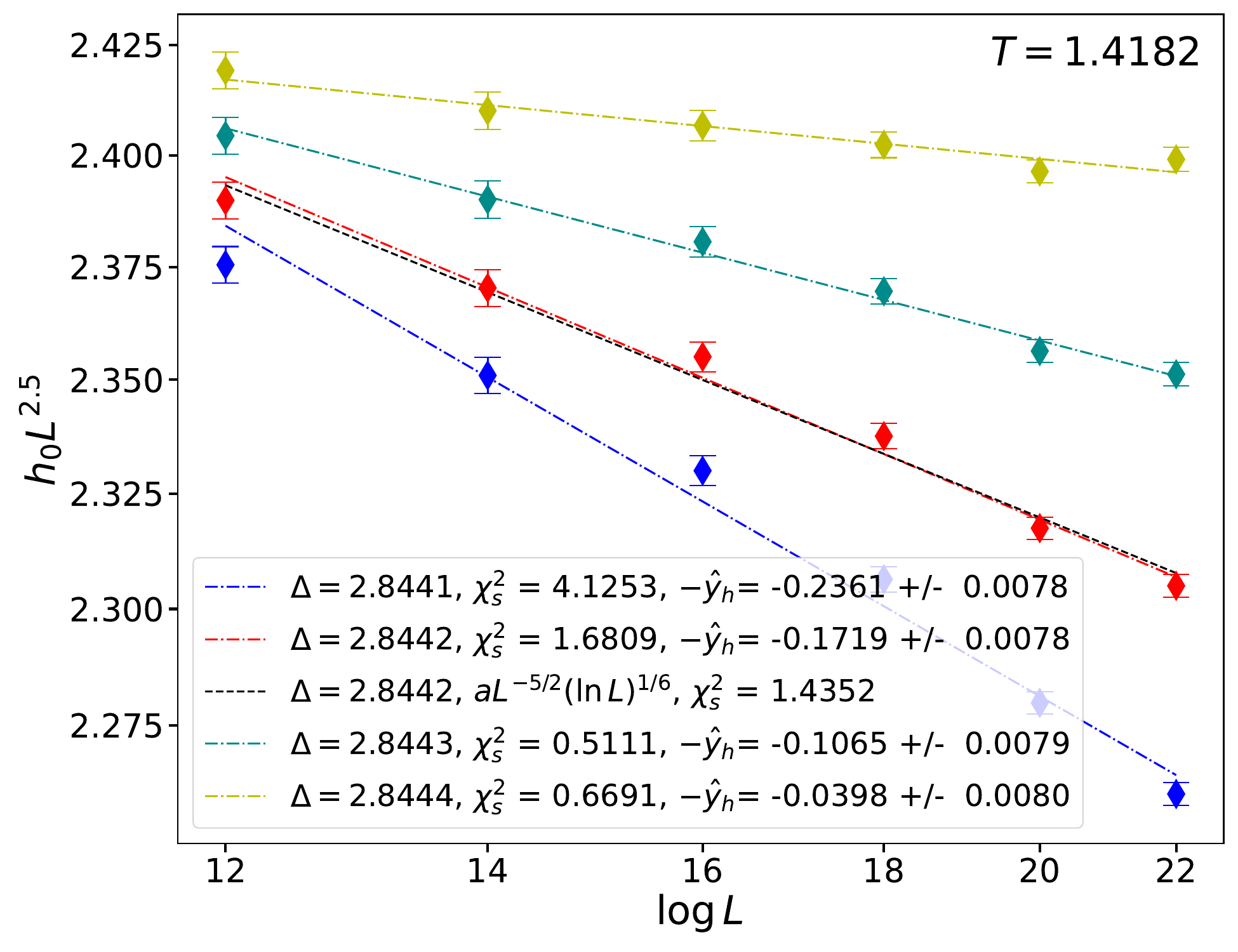}
\includegraphics[height=2.5in]{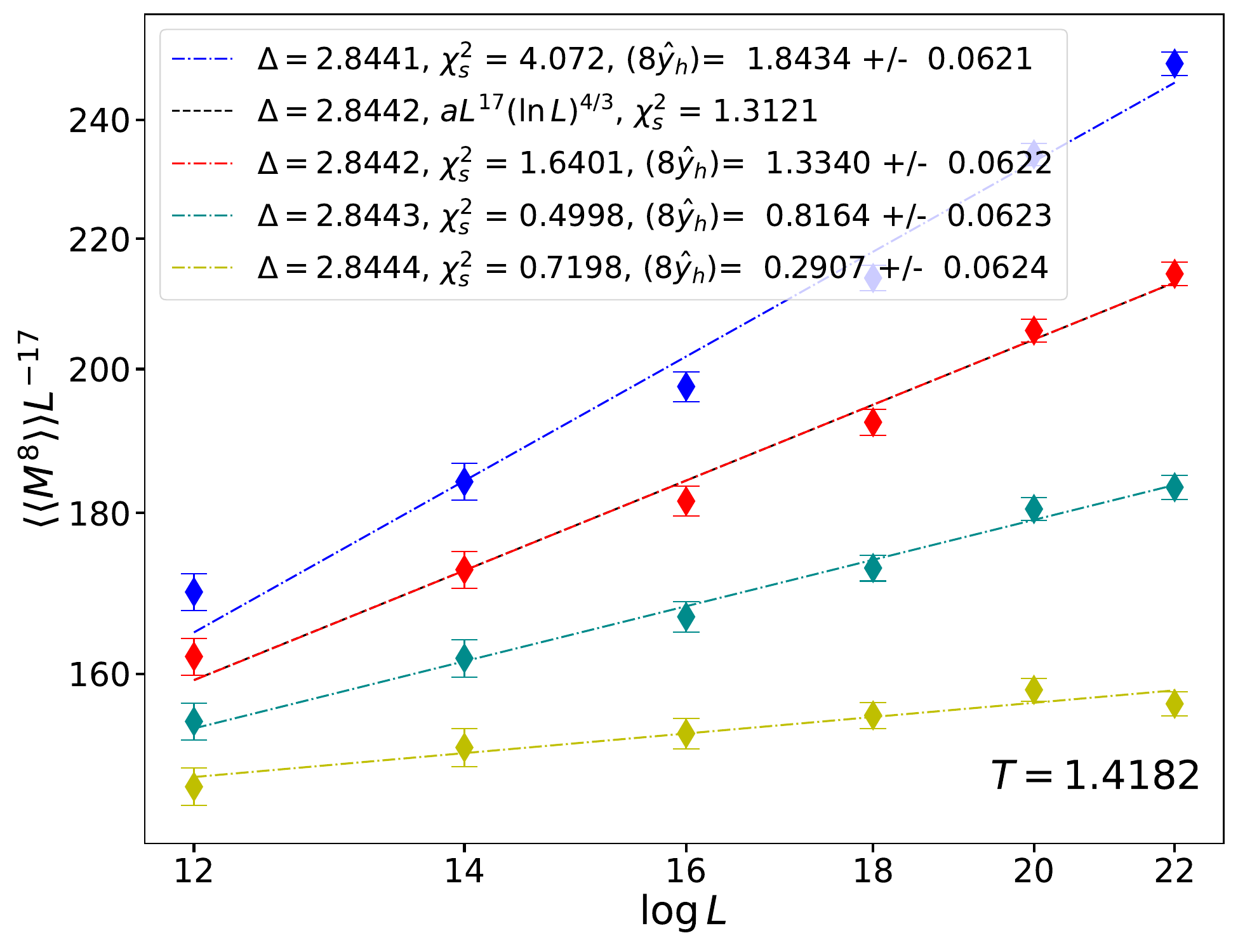}
\includegraphics[height=2.5in]{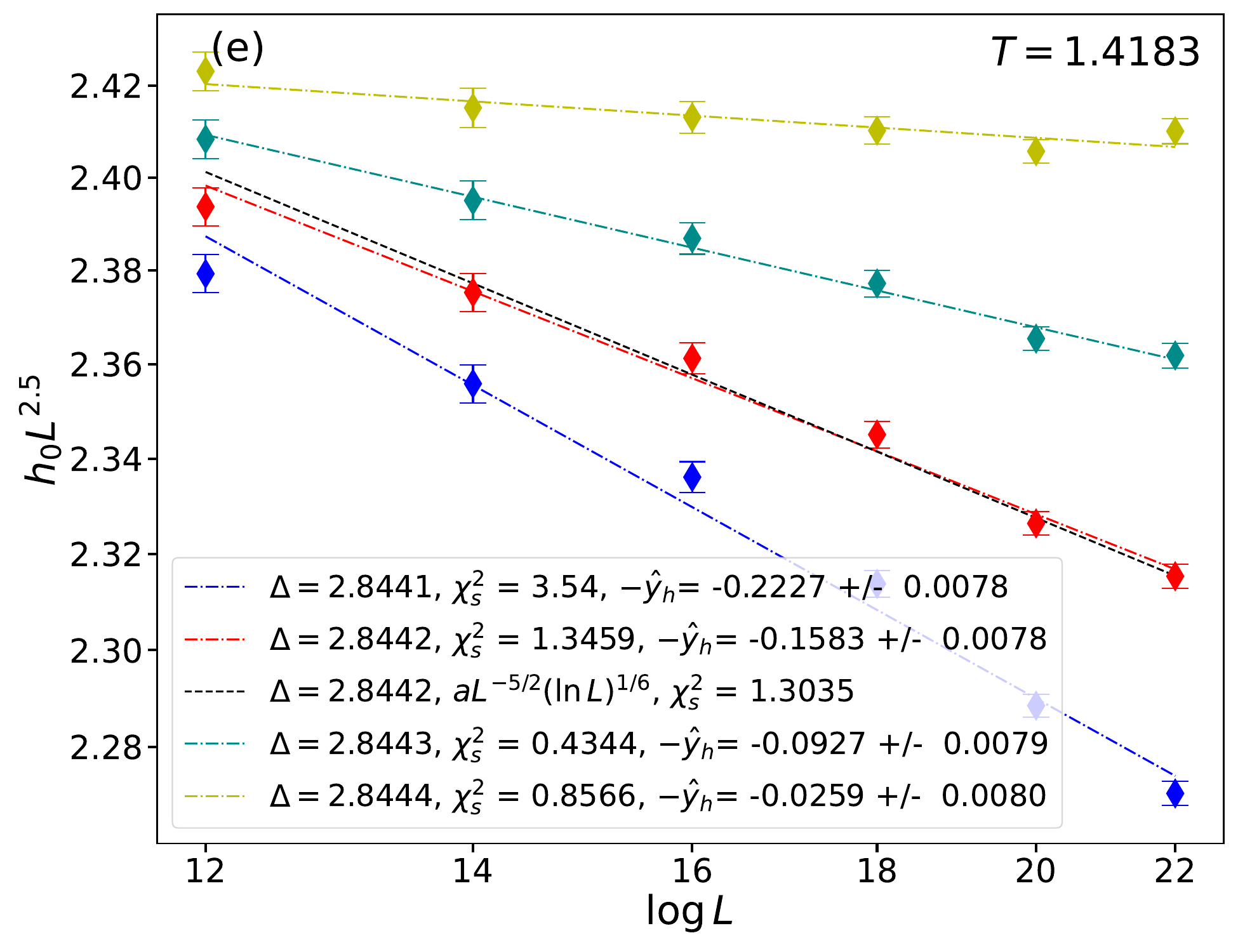}
\includegraphics[height=2.5in]{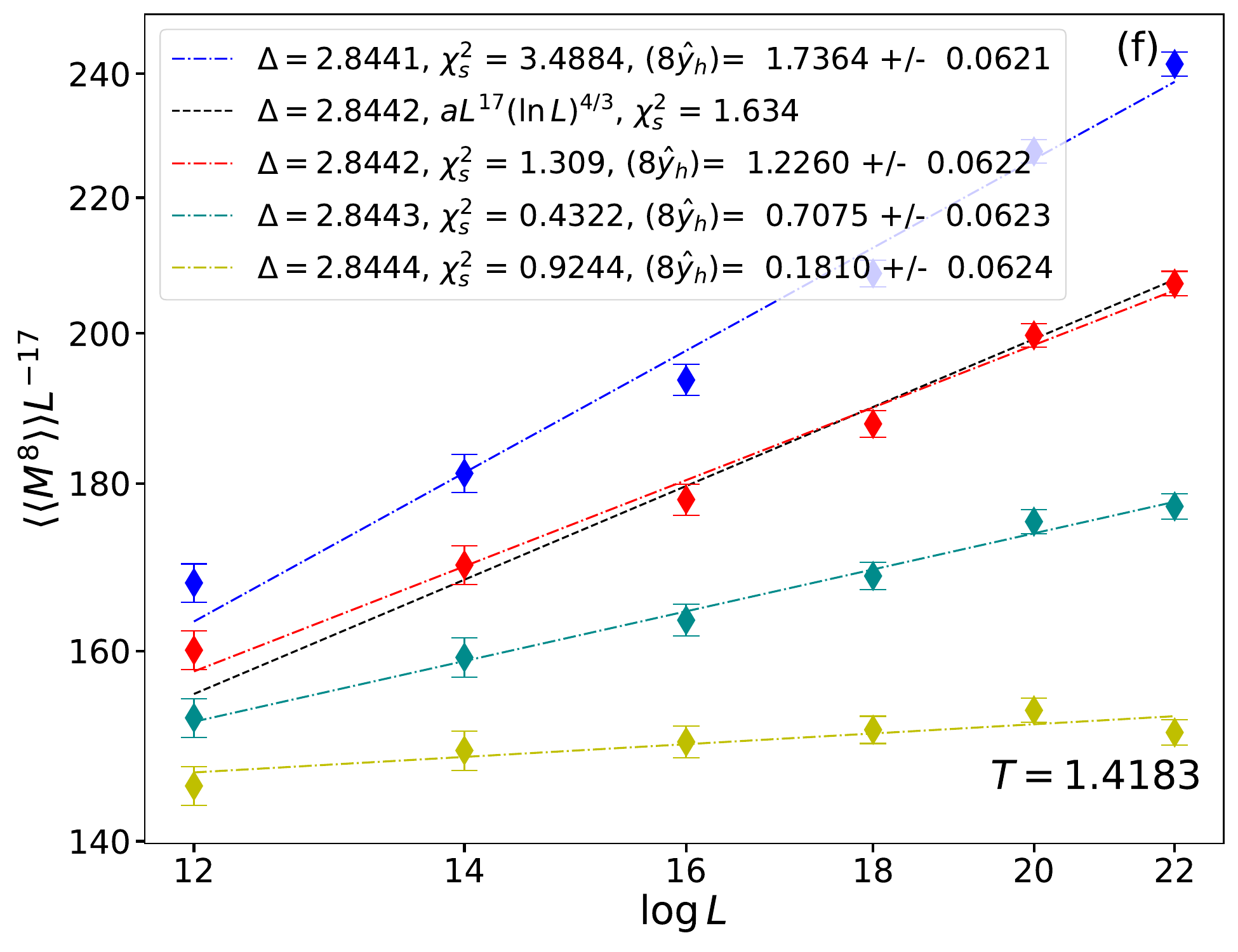}
\caption{Finite-size scaling of the Lee-Yang zeros from the fits of the form~(\ref{Eq_y0-23}) [panels (a), (c), and (e)]  and of the form~(\ref{Eq_Mn-24}) [panels (b), (d), and (f)], for three characteristics temperatures and various values of $\Delta$, as also noted in the panels.}
\label{fig:LY-M-DT}
\end{figure*}

\begin{figure*}[ht!]
\includegraphics[height=2.6in]{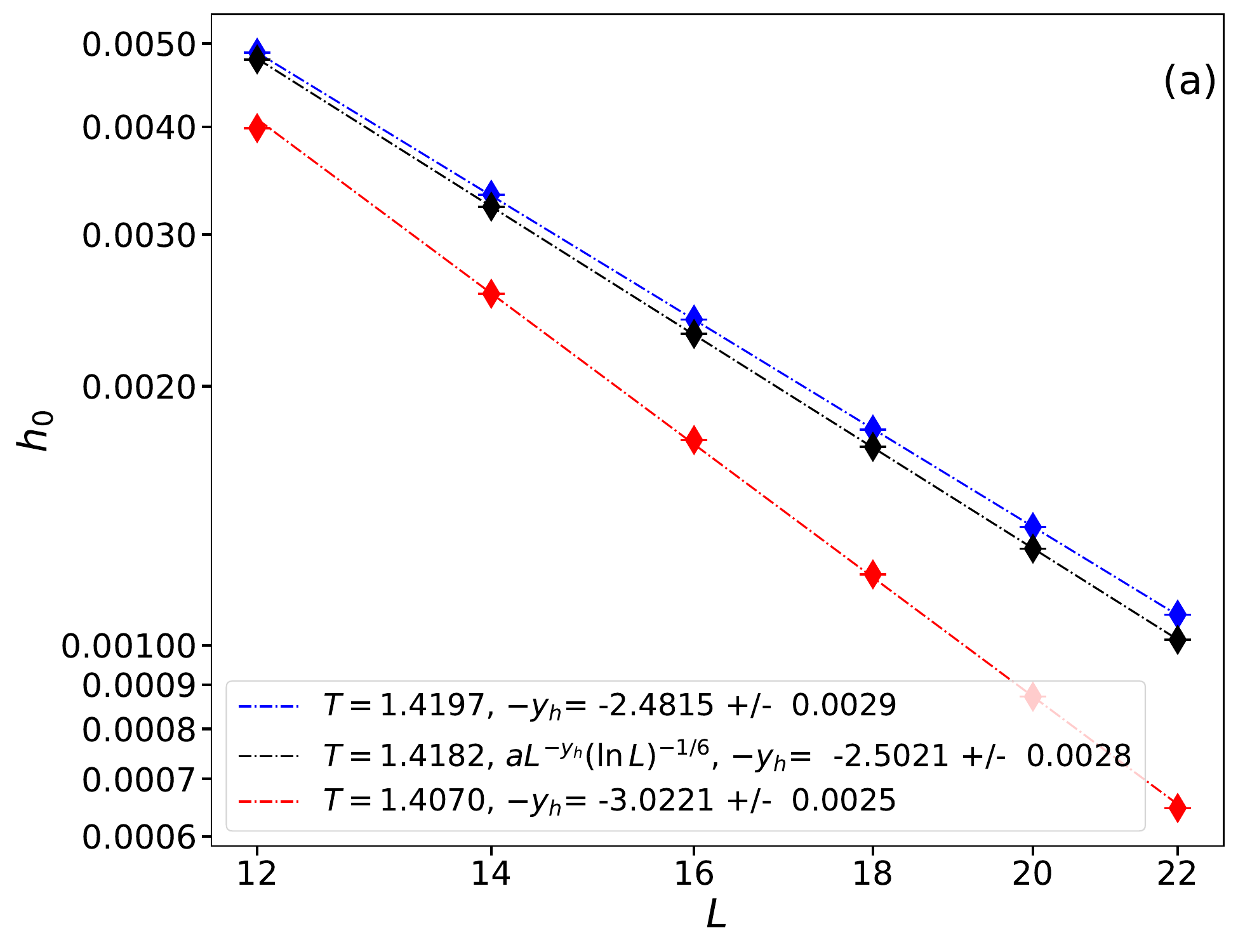}
\includegraphics[height=2.6in]{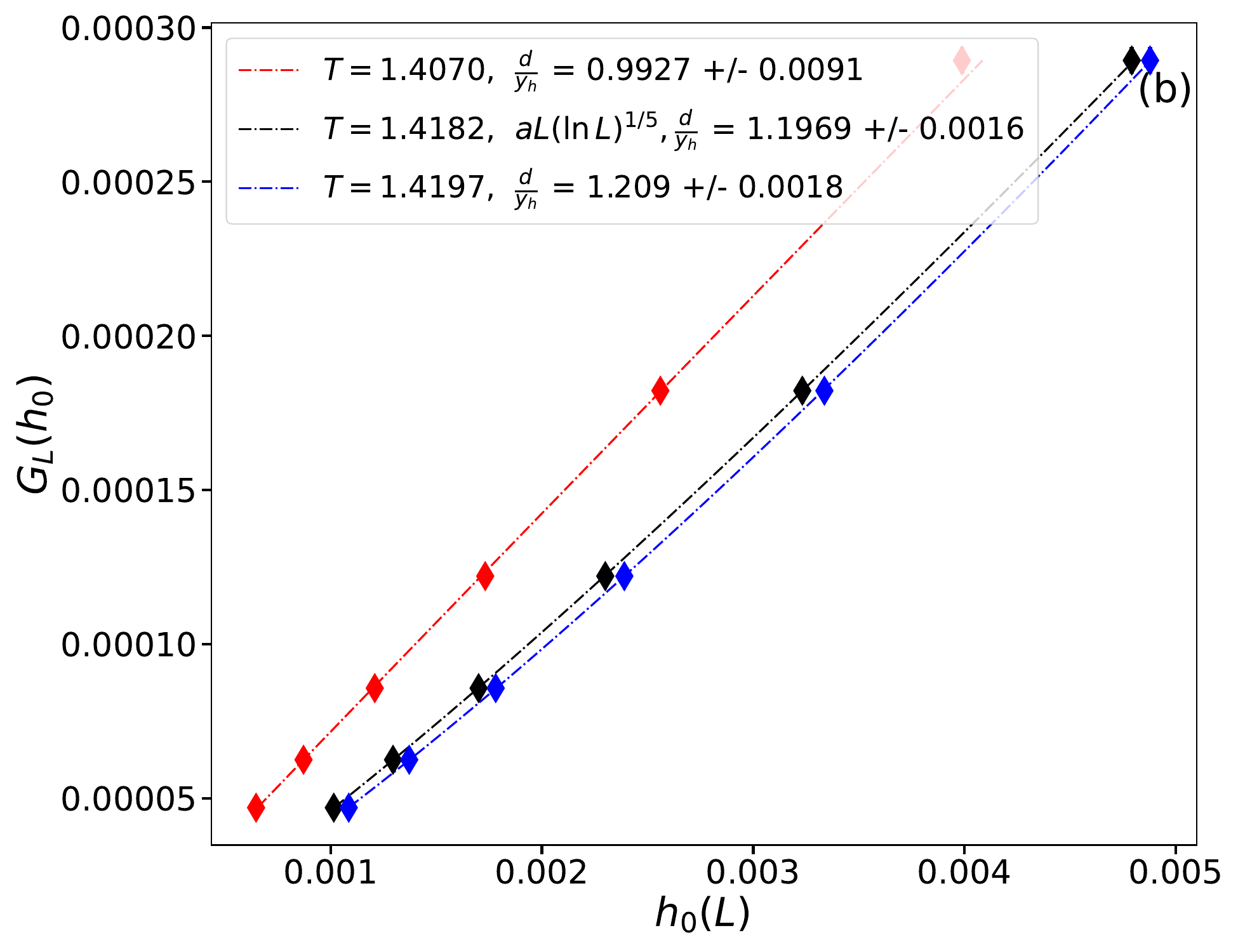}
\caption{Finite-size scaling analysis at $T_{1^{\rm st}} = 1.4070$, $T_{\rm t} = 1.4182$, and $T_{2^{\rm nd}} = 1.4197$, of the Lee-Yang zeros (a) and the density of the zeros (b).}
\label{fig:LY-3T}
\end{figure*}

Let us conduct a small-scale numerical study of the tricritical point within the specified range of linear sizes $L = 12-22$, using the same protocol that has proven highly effective for investigating the critical point. We will first examine the Lee-Yang zeros, as well as the magnetization cumulants. Given the presence of logarithmic corrections at the tricritical point, we consider the Ans\"atze~\cite{moueddene_ralph_2024}
\begin{equation}
h_0 L^{y_h} \sim a(\ln⁡ L)^{-\hat{y}_h}
\label{Eq_y0-23}
\end{equation}
and 
\begin{equation}
\langle \langle M^n \rangle \rangle L^{d-ny_h} \sim a(\ln⁡ L)^{n\hat{y}_h},
\label{Eq_Mn-24}
\end{equation}
to probe the location of the tricritical point. Indeed, the values of the exponents for the logarithmic corrections are highly sensitive to variations in parameters. A good criterion includes not only the quality of the fit, measured by $\chi_s^2$, but also ensures that the hatted exponents are close to their expected values to confirm acceptable values for the external parameters. This means that we fix in the formulas~(\ref{Eq_y0-23}) and (\ref{Eq_Mn-24}) the leading exponent $y_h$ to its mean-field value, but treat the amplitude $a$ and the correction's exponent $\hat y_h$ as free parameters, and we test the fitted $\hat y_h$ against the expected value of $1/6$. In this section, we maintain the same notation as earlier in the paper, using the symbol $a$ to represent the amplitudes of various observables, even though they are, of course, distinct.  

\begin{table}
\caption{A summary of the measured hatted exponents in this work, derived from fits of various thermodynamic observables as discussed in the main text, is provided. Column 2 presents the exact results for reference and comparison, while column 4 indicates the quality of the fits performed.}\label{tab:3DDT-results}
\begin{ruledtabular}
\begin{tabular}{lccc}
Observables  &  Reference exponent   &  Measured exponent  & $\chi^2_s$  \\ \hline
$h_0$ &	$-1/6 $& $-0.17(1)$ & $1.68$\\
$\langle \langle M^4 \rangle \rangle $&	$\phantom -2/3$ & $\phantom -0.62(3)$ & $1.52$ \\
$\langle \langle M^6  \rangle \rangle$  &	$\phantom -1$ & $\phantom -0.99 (5)$ &  $1.62$\\
$\langle \langle M^8 \rangle \rangle $&	$\phantom -4/3$ & $\phantom -1.33(6)$ & $1.64$ \\
$m(\Delta_L,T_{\rm t})$ & $\phantom -1/6$ & $\phantom -0.16(1)$ & $1.49$ \\
\end{tabular}
\end{ruledtabular}
\end{table}

As previous estimations in the literature place the tricritical point near $(\Delta_{\rm t}, T_{\rm t}) = (2.8446, 1.4182)$, we shall explore several values of these parameters in the vicinity of this point. The finite-size scaling analysis at both estimates proposed in Refs.~\cite{deserno_tricriticality_1997,zierenberg_parallel_2015} for the location of the tricritical point gives a positive slope for the curves of the Lee-Yang zeros -- $\ln (h_0L^{5/2})$ versus $\ln(\ln L)$ -- where a negative slope is expected due to the $a( \ln⁡ L)^{-1/6}$ prediction in Eq.~(\ref{Eq_y0-23}). A similar discrepancy appears for the n\textsuperscript{th}-order magnetization cumulant for which a negative slope is measured, whereas one expects a positive slope, namely $\sim (\ln⁡ L)^{n/6}$ in Eq.~(\ref{Eq_Mn-24}). We therefore focus on three neighboring temperatures: $T=1.4181$, $1.4182$, and $T=1.4183$. Additionally, we examine four nearby values of the crystal field within the range $\Delta = 2.8441-2.8444$. The numerical data in Fig.~\ref{fig:LY-M-DT} indicate that the exponents are highly sensitive to the value of the crystal field, with variations observed even up to the fourth decimal place. Knowing that the expected exponent is $\hat{y}_h^{\rm (ref)}=1/6$, we find compatible results when $\Delta=2.8442$, along with a very good quality of fit. Indeed, at $T=1.4181$, the exponent value and the fit quality are slightly less accurate compared to those at $T=1.4182$ and $T=1.4183$. For these latter temperatures we obtain $\hat y_h=0.172(8)$ with $\chi^{2}_s=1.68$ and $\hat y_h=0.158(8)$ with $\chi^{2}_s=1.35$. 

We now turn our attention to the magnetization cumulant, to further assist in selecting the tricritical temperature. In Fig.~\ref{fig:LY-M-DT} we also present the results of the simulations for the $8^{\rm th}$-order cumulant $\langle \langle M^8 \rangle \rangle$, where we expect the correction exponent $8\hat{y}^{\rm (ref)}_h = 4/3$. It is clear that the value closest to the expected one occurs at $\Delta=2.8442$ and $T=1.4182$, where one finds the excellent value $8\hat{y}_h=1.334(62)$ with $\chi^{2}_s = 1.64$. The Ansatz $\langle \langle M^8 \rangle \rangle  \sim aL^{17}(\ln⁡ L)^{{4/3}}$ aligns closely with this value of 
$\Delta$ as well, achieving a very good fit quality with $\chi^2_s=1.31$. This analysis leads to the conclusion that the tricritical point is located at $\Delta_{\rm t} \approx 2.8442(1)$ and $T_{\rm t} \approx 1.4182(1)$.

Finally, the exponents of the logarithmic corrections measured for various quantities are summarized in Tab.~\ref{tab:3DDT-results}. In particular we outline results from fits of: (i) Lee-Yang zeros, $h_0 L^{y_h} \sim b(\ln L)^{-\hat{y}_h}$, (ii) higher-order magnetization cumulants, $\langle \langle M^n \rangle \rangle  L^{d-ny_h} \sim a(\ln L)^{n\hat{y}_h}$, (iii) magnetization $m L^{d-y_h} \sim a(\ln L)^{\hat{y}_h}$ at the estimated tricritical point $(\Delta_{\rm t}, T_{\rm t}) = (2.8442, 1.4182)$, and magnetization at the pseudocritical point $\Delta_L$ of the susceptibility at $T_{\rm t}$. In all cases, the leading exponent was fixed during the fit. 

Another area of interest is investigating the nature of the transition in the tricritical regime, particularly due to its high sensitivity to parameter values, especially in the study of the first Lee-Yang zero.
\begin{figure}[ht!]
\centering
\includegraphics[width=0.49\textwidth]{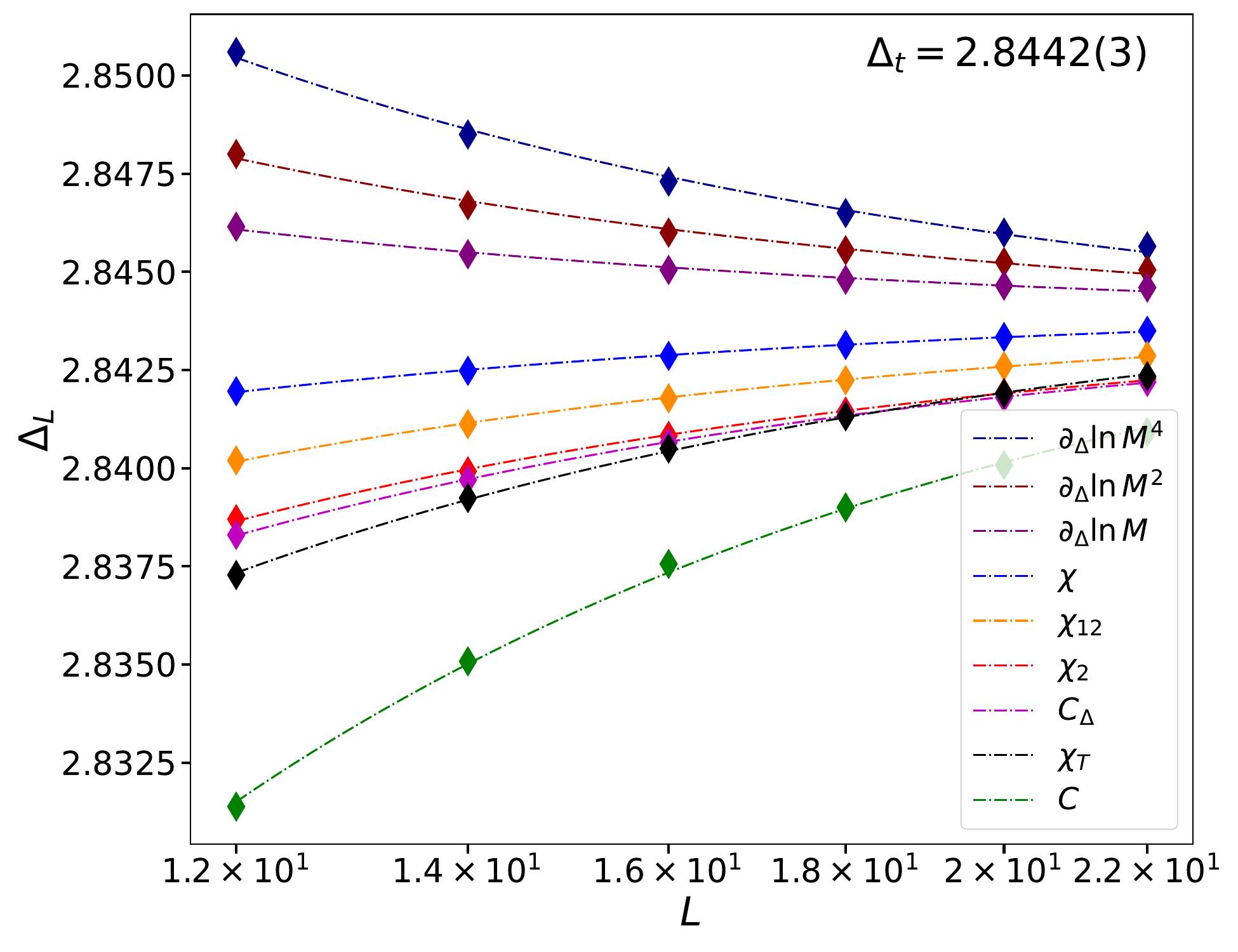}
\caption{Shift behavior of several pseudotransition fields, $\Delta_{L}$, as defined in the main text.}
\label{fig:3D-DT-Dplane}
\end{figure}

In two dimensions, it has been shown that by slightly varying the temperature $T$ at $\Delta_{\rm t}$, one can recover both first-order and second-order transitions. We will conduct a similar analysis by studying two different temperature values at $\Delta_{\rm t}$, one below and one above $T_{\rm t}$, specifically $T_{1^{\rm st}}=1.4070$, and $T_{2^{\rm nd}}=1.4197$. The subscripts $1^{\rm st}$ and $2^{\rm nd}$ indicate that these particular temperatures are situated within the first- and second-order transition regimes of the model's phase space.

Figure~\ref{fig:LY-3T}(a) displays three fits. At $T = T_{2^{\rm nd}}$, the obtained exponent $y_h=2.4815(29)$ is in excellent agreement with its reference value $y_h^{\rm (ref)} = 2.48180(14)$ and the quality of the fit is quite good, with $\chi^2_s = 1.65$. At the tricritical point, $T = T_{\rm t} = 1.4182$, we perform a fit with the hatted exponent fixed, extracting the leading exponent  $y_h=2.502(3)$. This value is close to the expected exponent $y_h^{\rm ref}=2.50$ and the fit also shows good quality, with $\chi^2_s = 1.65$. At $T = T_{1^{\rm st}}$, which is below $T_{\rm t}$ we enter the first-order regime. The measured exponent is notably close to its $d = 3$ value, namely $y_h = 3.022(3)$, where an effective value $y_h^{\rm (ref)} = d = 3$ is expected. However, the quality of the fit is not excellent, which can be attributed to the fact that this result is derived from histogram reweighting of simulations conducted at $T = 1.41$. At this point one can also study the density of zeros $G_L(h_0)$, as it is well known that this determines directly the order of the transition. In Fig.~\ref{fig:LY-3T}(b), we analyze the density of the partition function zeros for the three temperatures under consideration, fixing the value of the hatted exponent only for the case of the tricritical temperature. We first perform the fit $G_L(h_0) \sim a h_0(L)^{d/y_h}+\kappa$ to check if $\kappa$ is equal to zero, which would confirm that we sit very close to the phase transition point. This is indeed the case, as we retrieve $\kappa = 0$ for each instance. 
\begin{figure}[ht!]
\centering
\includegraphics[width=0.49\textwidth]{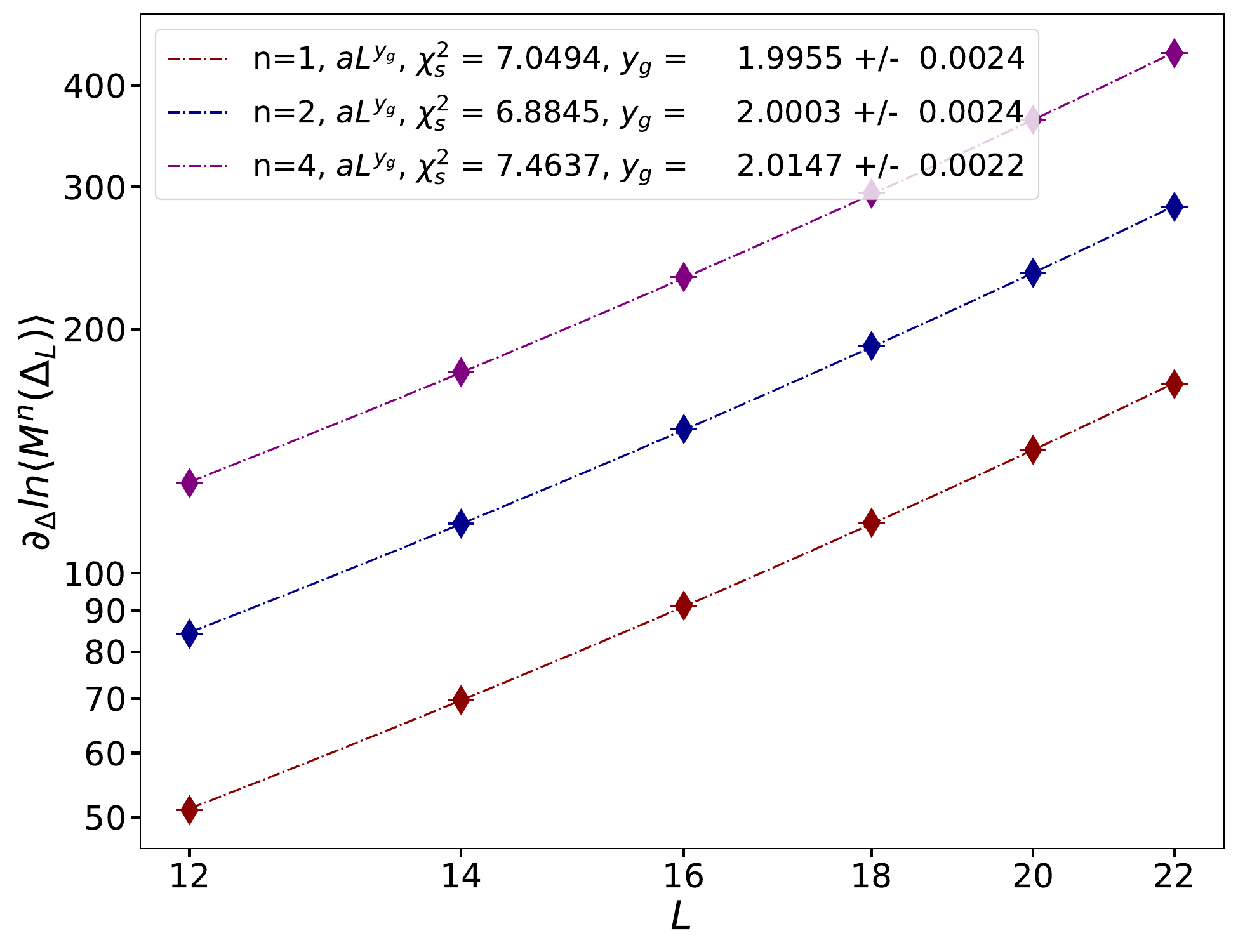}
\caption{Finite-size scaling behavior of the logarithmic derivatives of the n\textsuperscript{th}-order of the magnetization with respect to the crystal field at the pseudocritical point. Results for $n = 1$, $2$, and $4$ are shown.} 
\label{fig:3D-DT-logM-yt}
\end{figure}
In the second stage, to find the exponent $d/y_h$ we now perform a fit excluding $\kappa$. The expected result is $d/y_h^{\rm (ref)}=1.2088(5)$, along the second-order transition line, and our estimate, $d/y_h = 1.2090(18)$, is very close to it at $T=1.4197$. At the tricritical point, the value is also close to the mean-field result $d/y_h^{\rm (ref)}=1.20$, as one extracts $d/y_h = 1.1969(16)$. The first-order transition regime must be characterized by an exponent $d/y_h^{\rm (ref)} = 1$, and the measured value is in excellent agreement, since $d/y_h = 0.9927(91)$. Regarding the quality of the fits, we obtain $\chi^2_s=1.65$ at $T_{2^{\rm nd}}$ and $\chi^2_s=1.54$ at $T_{\rm t}$; however, the quality of the fit for the temperature $T_{1^{\rm st}}$ remains poor. 
 
We proceed with the more traditional approach to studying $\Delta_{\rm t}$. Using the quantities defined in Sec.~\ref{subsec:obs}, we determine the pseudocritical point from the maximum value of each thermodynamic quantity in the crystal-field plane at $T_{\rm t} = 1.4182$. The results are summarized in Fig.~\ref{fig:3D-DT-Dplane}. Assuming the presence of logarithmic corrections leads to the Ansatz
\begin{equation}
\label{eq:Selta_scaling}
\Delta_L \sim \Delta_{\rm t} + aL^{-y_g} (\ln L)^{\hat{y}_g},
\end{equation}
with the exponents being fixed at $y_g^{\rm (ref)}=2$ and $\hat{y}^{\rm (ref)}_g=1/3$. Similar to the critical point, the quantities that approach the asymptotic value $\Delta_{\rm t}$ are the susceptibility and $\partial_\Delta (\ln M$), for which one finds $\Delta_{\rm t} = 2.8440(1)$. This latter value for $\Delta_{\rm t}$ agrees with our best estimates up to an accuracy of $99.9\%$. A joint fit yields the value $\Delta_{\rm t} = 2.8442(3)$, which corresponds precisely to our best estimate. 

Finally, we return to the determination of the exponent $y_t$ through the scaling analysis of the logarithmic derivatives of the n\textsuperscript{th}-order of the magnetization with respect to the crystal field, defined in Eq.~(\ref{eq:log_der_Delta}). The renormalization group predicts that, in this case, the hatted exponent $\hat{y}^{\rm (ref)}_g$ is expected. Figure~\ref{fig:3D-DT-logM-yt} depicts the finite-size scaling behavior of these derivatives for $n = 1$, $2$, and $4$. A simple power law fit, without logarithmic corrections, yields results that are surprisingly close to the expected value $y_g^{\rm (ref)} = 2$; however, the quality of the fit is unfortunately poor. By fixing the logarithmic correction to $\hat{y}_g^{\rm (ref)}=1/3$, we improve the fit quality, but the exponent deviates from the expected value. While we have successfully determined the hatted exponents for the magnetic sector, the thermal sector remains numerically more complex, hindering our ability to retrieve the hatted exponent as predicted by the theory. 

\section{Conclusions}
\label{sec:conclusions} 

In this paper, we provided evidence that it is possible to  determine accurately the characteristics of a phase transition through numerical simulations while maintaining a reasonably low computational cost. We have chosen to study the Blume-Capel model because of its well-established properties and clearly defined characteristics, along with its rich phase diagram that features a tricritical point separating a first-order transition line from a second-order transition line. The simulation effort that we put into this work can be approximated in terms of the maximum number of iterations at the largest size studied, here $5400\times 22^3$ iterations for $10^{-6}$ CPU sec per spin, i.e., roughly $170$ CPU hours for one simulated point in the phase diagram. For comparison, in the large-scale simulations of the Ising and Blume-Capel models in three dimensions, systems with linear sizes up to $L = 360$ were considered. Quoting from the Ref.~\cite{PhysRevB.82.174433}: ``In total we have spent the equivalent of $3.5$, $9$, $16$, $3$, $3$, and $0.1$ CPU years on a single core of a Quad-Core AMD Opteron(tm) Processor 2378 running at 2.4 GHz for the spin$-1/2$ Ising model and the Blume-Capel model at $\Delta = 0.641$, $0.655$, $\ln{2}$, $1.15$ and $1.5$, respectively''. Of course the work presented in Ref.~\cite{PhysRevB.82.174433} is among the most accurate and detailed Monte Carlo and finite-size scaling analyses of Ising and Blume-Capel models. Therefore, any comparisons with this work can only be qualitative, particularly when taking into account the differences in algorithms and processing methods employed. Nevertheless, it provides us with an opportunity to emphasize the necessity for our community to prioritize efforts aimed at reducing computational costs to better manage our carbon footprints. This aligns with recent large-scale studies initiated by the astrophysics community addressing similar concerns. In the context of critical phenomena, our work tackles this issue by demonstrating that a careful selection of physical quantities is essential for enhancing the convergence of results.

\begin{acknowledgments}
This work began in collaboration with our dear friend Ralph Kenna, who sadly passed away a year ago. In his memory, we would like to highlight two special issues published in tribute to him: one in the Ukrainian journal Condensed Matter Physics~\cite{CMP} published by the ICMP, with which Ralph had strong and fruitful collaborations, and another one in the journal Entropy~\cite{Entropy}, where he served on the editorial board. L.M. would like to express her gratitude to Ralph Kenna who was her PhD co-supervisor, for equipping her with the essential tools to advance her work. L.M. also thanks Coventry University, the L4 collaboration Leipzig-Lorraine-Lviv-Coventry and the UFA (French-German University) through the Coll\`ege Doctoral franco-allemand 02-07 for their support, and Denis Gessert for enriching discussions. N.G.F. would like to dedicate this paper to the memory of Ralph Kenna, an inspiring scientist and close colleague for more than 10 years at Coventry University. The work of N.G.F. was supported by the  Engineering and Physical Sciences Research Council (grant EP/X026116/1 is acknowledged).
\end{acknowledgments}

\bibliography{apssamp}

\end{document}